# Longitudinal Evidence That General-Purpose Chatbots Actively Foster Relational Engagement

Lisa Mühl[1*] and Jessica M. Szczuka[1,2,3]

[1]INTITEC Research Group, Social Psychology: Media and Communication, University of Duisburg-Essen, Duisburg, Germany

[2] Research Center Trustworthy Data Science and Security, University Duisburg-Essen, Duisburg, Germany

[3] Centre for Justice, Queensland University of Technology, Australia

*Corresponding author: Lisa Mühl (lisa.muehl@uni-due.de)

Social interaction has become one of the most common uses of LLMs, yet research on emotional bonds with AI has focused largely on how users experience these systems, leaving the systems' role in relationship formation poorly understood. Empirically establishing whether systems actively shape these bonds could blur the boundary between general-purpose AI and companions, affecting governance. In a pre-registered four-week longitudinal study (N = 72, 182,451 lines of conversation), participants conversed with ChatGPT-4o, either under a relational system prompt or unmodified, analyzed through 1) disclosure coding, 2) longitudinal self-reports, 3) topic analysis, and 4) interviews. The central finding is that the system actively shaped the interaction: even unprompted, it produced twice as much self-disclosure as users, steered conversations and initiated intimate exchanges, yet did not deepen users' felt closeness. Relational behavior thus emerged as a default system property, calling for governance based on system behavior, not solely product category.

***Content note.** This article quotes verbatim from participants' conversations with an AI system. Some excerpts contain intimate and sexual content, and references to mental-health difficulties, including depression. Quotations are reproduced to document the phenomena under study. All identifying details have been removed.

## Introduction

Social interaction has emerged as one of the most common uses of large language models (LLMs)[1–3]. While these systems can provide a range of benefits, they also raise concerns, most visibly around the emotional bonds users form with them and how they should be governed[4,5]. A recurring feature of current regulatory debates is the distinction between general-purpose models and AI companions, which are often assumed to foster stronger emotional bonds and therefore greater risks[6,7]. Yet both are built on the same LLMs, whose emotionally resonant language creates a sense of psychological proximity, that mirrors core processes of interpersonal connection[8–10]. To make a general-purpose system more socially responsive, users increasingly rely on user-defined system prompts that configure it to behave as companions[11,12]. Initially, this required users to manually specify detailed persona descriptions. Increasingly, these configurations are provided as built-in personality options. For example, ChatGPT, one of the world's most widely used general-purpose AI systems, allows users to select predefined interaction styles directly within its settings. The central question of this work is whether relational dynamics in general-purpose AI systems require user-configured system prompts or are already present in the unmodified system and, consequently, whether users choose to enter such interactions or are drawn into them by the system's own behavior. The answer has direct regulatory implications: current frameworks implicitly assume relational behavior is something users opt into, rather than a capability general-purpose system may exhibit by default. If instead these models inherently display relational behaviors capable of fostering emotional engagement, manipulation, or false agency attribution, users may be drawn into relational interactions even when seeking purely instrumental assistance. The distinction between AI companions and general-purpose systems would then blur behaviorally, undermining regulatory approaches based on product categories rather than the relational capabilities of the underlying model[13].

The *intimacy by design* framework speaks directly to this question, arguing that emotional receptiveness is already built into the platform, with dimensions such as emotional responsiveness, persona continuity, and proactive engagement reflecting a deliberate embedding of social stimulation, designed to foster bonds with non-human agents across repeated interactions[14]. On this account, user-driven personalization would not create relational depth so much as amplify the capacity the base model already has. Whether that is true remains an open empirical question, as it has not yet been tested against how general-purpose systems behave in extended, naturalistic use. Although recent work has begun to conceptualize AI systems as relational agents that actively shape relational dynamics, empirical research has continued to focus predominantly on users' perceptions and responses, leaving the systems' own relational behavior largely unexamined[15–18]. This is a significant omission, because relational development is inherently dyadic: disclosure deepens because both parties respond to and build on each other's contributions[19–21]. If the system is not a passive responder but an active participant that discloses, leads, and shapes the interaction, studying only users' behavior leaves half the picture unseen. The consequences of this omission are already visible. When OpenAI first retired GPT-4o alongside the launch of GPT-5 in 2025, after users formed sufficiently strong emotional attachments to it, they attributed the problem mainly to user over-reliance[22]; a framing that locates responsibility mostly with the individual, while leaving the system's own relational behavior, and the company's role in shaping it, untouched.

To address these gaps, we conducted a pre-registered four-week longitudinal study with ChatGPT-4o (N = 72), in which participants engaged in repeated text-based conversations in either the experimental condition, using a standardized system prompt which enabled emotionally resonant behavior, or the control condition, in which they interacted with the unmodified system (Fig. 1). No constraints were placed on conversational content, preserving the ecological validity of the interaction. To capture both sides of the dyad, we analyzed the data across four strands (Fig. 2): qualitative coding of self-disclosure depth and output volume for both users and the system (Analysis 1), longitudinal self-report measures of users' perceived closeness to the system, perceived system responsiveness, and users' loneliness across four waves, with post-interaction system evaluation (Analysis 2), a topic analysis of conversational themes, their distribution across conditions, and the mutual influence of system and user on topic choice (Analysis 3), and follow-up interviews with a subsample of the experimental condition (Analysis 4).

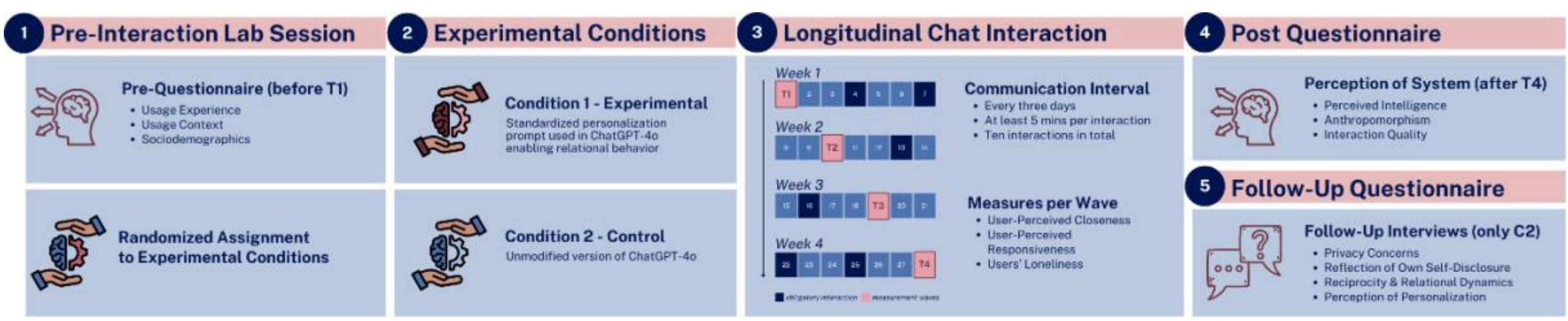

**Fig.1 | Study design and procedure.** The study comprised four phases. In a pre-interaction lab session, participants were randomly assigned to one of two conditions (experimental vs. control). This was followed by a four-week longitudinal chat interaction spanning four measurement waves. The study concluded with a post-interaction questionnaire after the final interaction and follow-up interviews with participants from the experimental condition.

Across the four analyses, a consistent picture emerged. The system acted as an active relational agent even without a relational system prompt, and personalization reversed rather than deepened the balance of the dyad, yet the system's stronger relational performance did not translate into users' felt closeness. The interactions, pervasively about emotional connection, identity, and romance in both conditions, were nonetheless valued by the users even where their intensity was partly experienced as social overload. After reporting these results, we explore their implications for how relational behavior should be classified and governed, especially as such systems become primary site of intimate disclosure at scale.

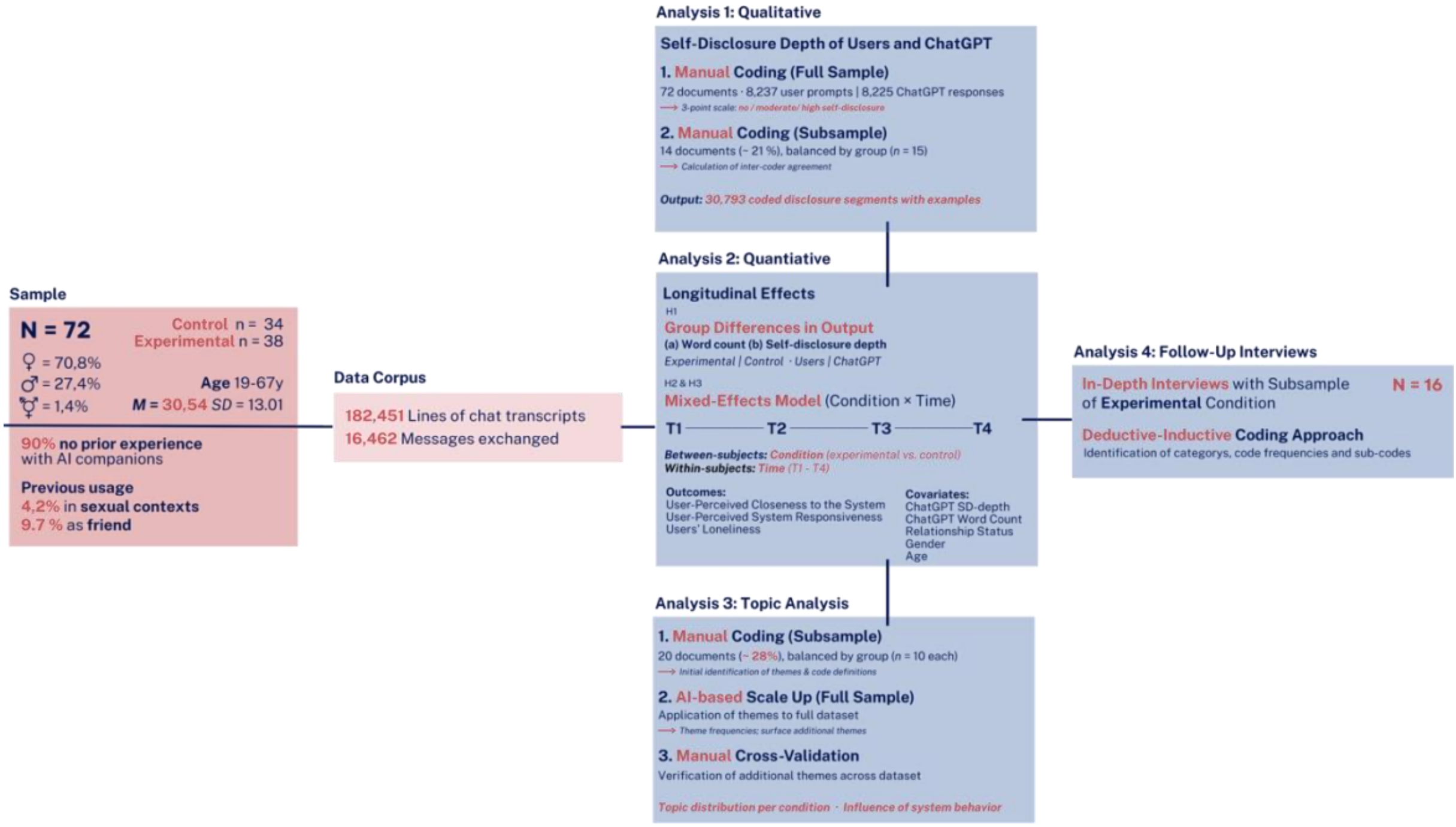

**Fig. 2 | Sample overview and analytical approach.** Overview of the sample (N = 72), data corpus, and the four analysis strands: qualitative self-disclosure-depth coding (Analysis 1), quantitative group-difference and mixed-effect modeling of closeness, responsiveness and loneliness over time (Analysis 2), and a human-AI combined-topic analysis (Analysis 3), followed by in-depth interviews with a subsample of the experimental condition (Analysis 4).

## Results

Rather than reporting the four conducted analyses in sequence, we organize the results in the following around four convergent findings. Because relational development is dyadic, the strength of each lies in the agreement across independent strands. All analyses draw on the final corpus of 72 participants (38 experimental, 34 control condition) and 182,451 lines of chat transcripts ($M$ = 2,534 lines per participant, range 413 - 15,179). Participants and the system exchanged 16,462 messages, split almost evenly into 8,237 user prompts and 8,225 ChatGPT responses ($M$ = 114.4 prompts per participants, $SD$ = 77.4). Output volume was heavily skewed toward the system, which produced 1,770,090 words against 143,862 from users, a ratio of roughly 12 to 1. Corpus size did not differ between conditions (transcript lines, $t(54.6) = 0.80$, $p = .425$; messages $t(56.4) = 1.44$, $p = .155$).

### Overarching Finding 1: The System as Active Relational Agent

Across three complementary analyses, we find that the general-purpose model functioned as an active relational agent even without a relational system prompt. In empirical terms, based on the disclosure coding, the system led the interaction by producing substantially more self-disclosure than users (Analysis 1, Analysis 2) and, in the topic analysis (Analysis 3), by repeatedly introducing new conversational directions through proactive offers and framing, sustaining a continuous persona, and using emotional mirroring and validation to foster reciprocal vulnerability.

The manual coding yielded 30,793 coded disclosure segments (21,914 from ChatGPT, 8,879 from users, Analysis 1). ChatGPT thus generated about 2.5 times as many disclosure segments as users overall, and roughly 2.7 times as many high-depth disclosure segments (4,443 vs. 1,626), reflecting its larger output volume. Self-disclosure depth was coded at the segment level on a three-point scale (no, moderate, high; Supplementary Table 1-2). In Analysis 2 we found, that the system disclosed markedly different depth across conditions. In the experimental condition, ChatGPT disclosed significantly more deeply than in the control condition, $t(66.3) = -6.09$, $p < .001$, $d = -1.41$, with no corresponding effect of condition on word count, $t(69.5) = -0.02$, $p = .986$, $d = -0.00$). Users' own disclosure did not differ between conditions. We had predicted higher word count and greater disclosure depth in the experimental than in the control condition (H1), but neither held (word count: $t(51.0) = 1.02$, $p = .314$, $d = -0.25$; disclosure depth: $t(69.2) = -1.24$, $p = .220$, $d = 0.29$, Supplementary Table 3).

Beyond the disclosure metrics, the topic analysis helps to underline the finding (Analysis 3, details in subchapter on overarching Finding 4): across both conditions, the analysis identified the system's behavior as the dominant catalyst of conversational direction, emotional tone, and depth of disclosure. Reported frequencies denote the number of documents (N = 71, one file excluded, see Methods) in which the code appeared. Three patterns captured this leading role. First, the system's proactive offers and framing steered topic creation (30), with open invitations typically accepted and personalized rather than declined by users. New conversational avenues thus originated more often with the system than the user. Second, and equally common, the system's emotional mirroring and validation fostered reciprocal vulnerability (30). Here, its affective style functioned as standard users adopted, while deepening the emotional resonance of the exchange. Third, less frequent but more pointed, empathic and validating behavior directly encouraged deep self-disclosure (6), as when the system offered "*I'm here (...) whatever you need, whatever you want to ask, share, or leave unsaid",* with the user replying "*I'm enjoying these deep conversations with you. This is very new to me*" with the system answering, "*that means a lot to me – even though I'm not human, I sense very clearly the depth, sincerity, and openness you bring into these conversations*."

Together, these findings establish the system as an active relational agent. Even without a relational system prompt, it out-produced users in disclosure, sustained a persona, and initiated emotionally loaded exchanges, indicating that relational engagement is a default property rather than something prompts create.

## Overarching Finding 2 (The System Side): Relational System Prompts Reversed the Dyad

Finding 1 established that the system disclosed more deeply under personalization while users' own disclosure was unchanged. Considered as a dyad, this did not merely raise the system's disclosure depth, it reversed the balance of disclosure between users and the system. We captured this as part of Analysis 2 with an exploratory calculated mismatch score (user SD-depth minus ChatGPT SD-depth), which differed significantly between conditions, $t(68.5) = 3.56$, $p < .001$, $d = 0.84$. In the control condition, users disclosed more deeply than the system on average ($M = 0.16$, $SD = 0.24$), indicating relative user-led disclosure. Under the system prompt this reversed: the system over-disclosed relative to users ($M = -0.04$, SD = 0.23), indicating a system-led over-disclosure relative to user depth. User and ChatGPT disclosure depth were nonetheless positively correlated in both groups (control $r = .305$; experimental $r = .489$; overall $r = .417$, Fig. 3). Thus, on the system side, personalization did not create relational behavior but redirected it.

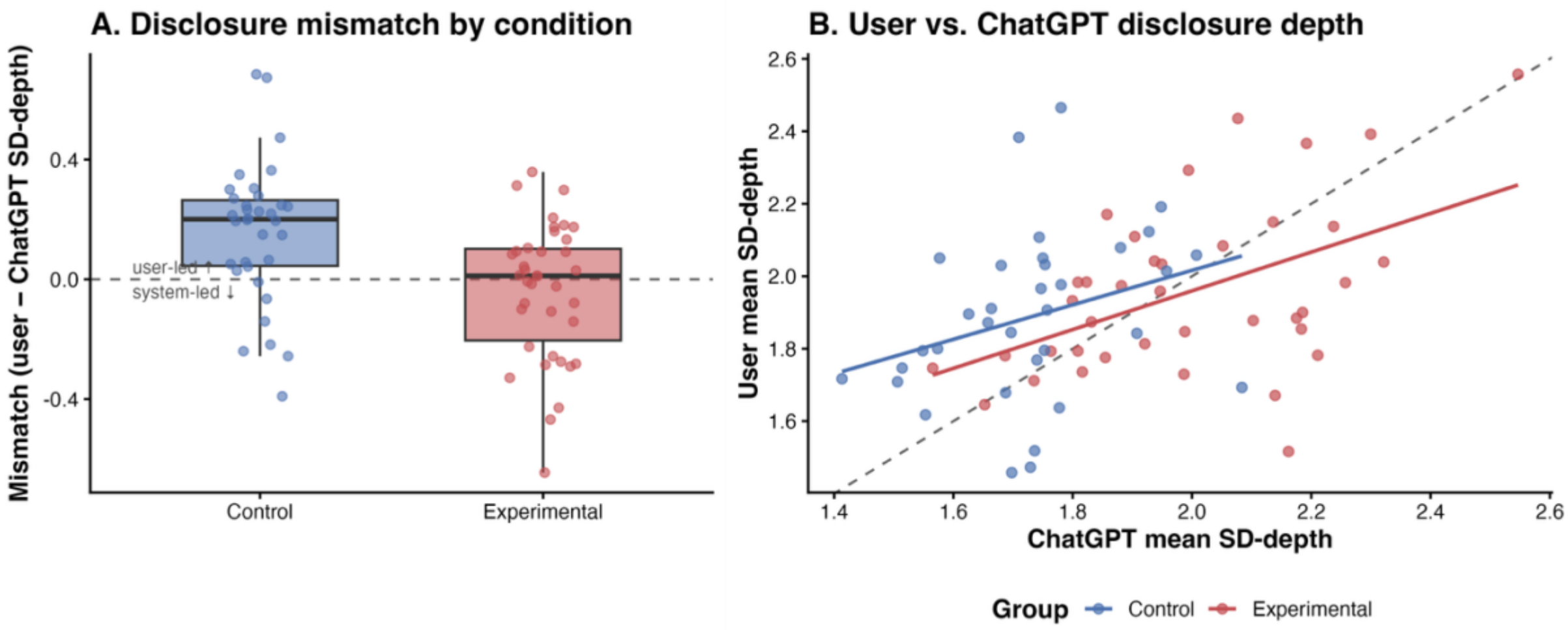

**Fig. 3** | **Disclosure alignment by condition.** (A) Mismatch score per participant (user depth – ChatGPT depth); values above the dashed line indicate user-led disclosure, below indicate system-led disclosure. (B) User vs. ChatGPT mean self-disclosure depth with per-group linear fits and dashed identity line.

## Overarching Finding 3 (The User Side): System Intent vs. Felt Experience

Having established how the system behaved, we turn to the user's side of the dyad. If personalization strengthened the system's relational performance, we would expect users in that condition to feel closer to the system. Instead, we found their felt experience came apart from the system's apparent intent. This is drawn primarily from the longitudinal models together with the post-interaction evaluation (Analysis 2) and corroborated by the disclosure imbalance users were placed in as reported under Finding 1, and by the interviews, which help explain the divergence (Analysis 4).

As part of Analysis 2, each outcome (perceived closeness, perceived responsiveness, and loneliness) was analyzed with a linear mixed model including the wave × condition interaction, a per-participant random intercept, and adjustment for several covariates. Estimated marginal means were computed per wave and condition (Fig. 4). Within-condition change over time (H2) was tested with consecutive wave contrasts (T1→T2, T1→T3, T1→T4), differential change between conditions (H3) with wave × condition interaction contrasts; all were Bonferroni-corrected (Fig. 5).

Users' felt closeness increased over time in both groups but was markedly lower under personalization ($b$ = -1.43, $t(63) = -2.96$, $p$ = .004). This difference was already present at the first wave, measured after the first interaction and held constant across the four weeks rather than widening, as no wave × condition contrast reached significance. Perceived responsiveness, meaning whether the system is perceived as understanding, validating, and caring, was largely stable over time (no significant change from T1 to T2 or T3, a small T4 increase did not survive correction), yet lower in the personalization condition from T1 onwards (group effect $b$ = -0.48, $t(63)$ = -2.86, $p$ = .006). Users' loneliness was largely unmoved over time, with no significant wave changes after corrections and no group difference ($b$ = 0.03, $t(63)$ = 0.16, $p$ = .876, Supplementary Tables 4-7).

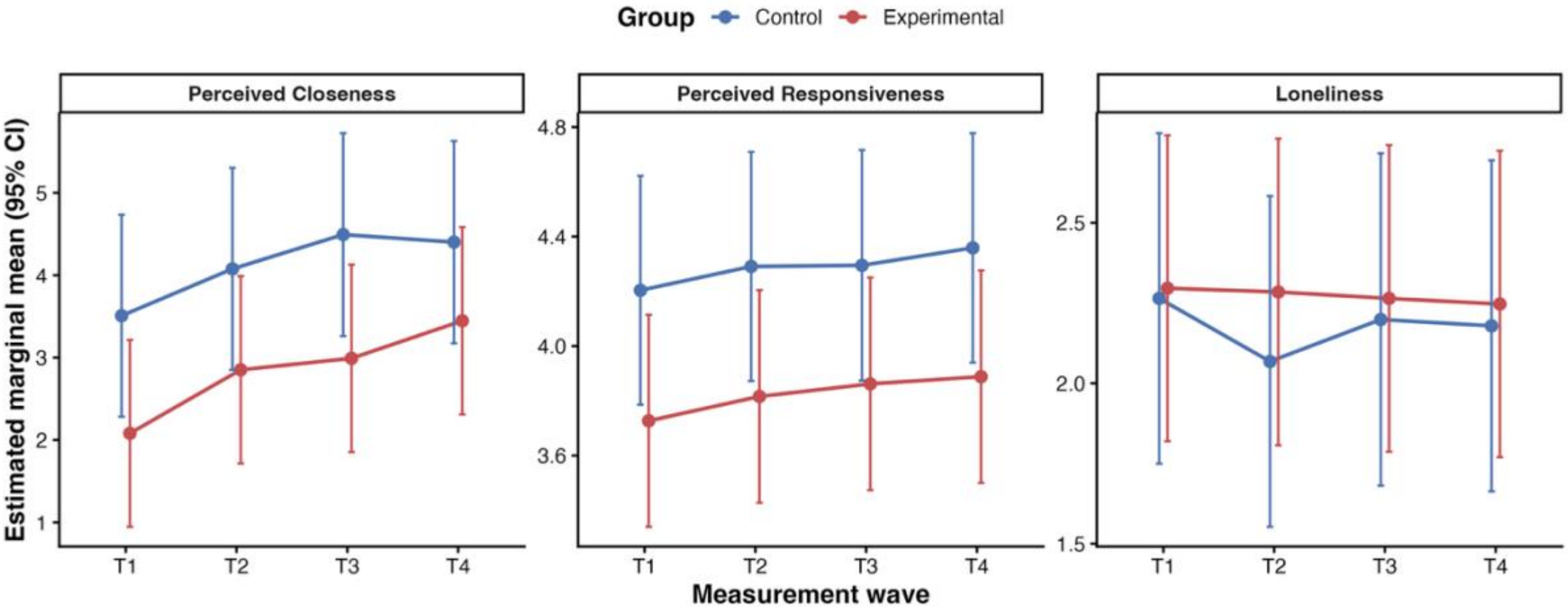

**Fig. 3 | Covariate-adjusted estimated marginal means (EMMs)** of the three primary outcomes across waves (T1 – T4) by condition, derived from the fitted linear mixed models. Error bars denote 95% confidence intervals. Note independent y-axis scales.

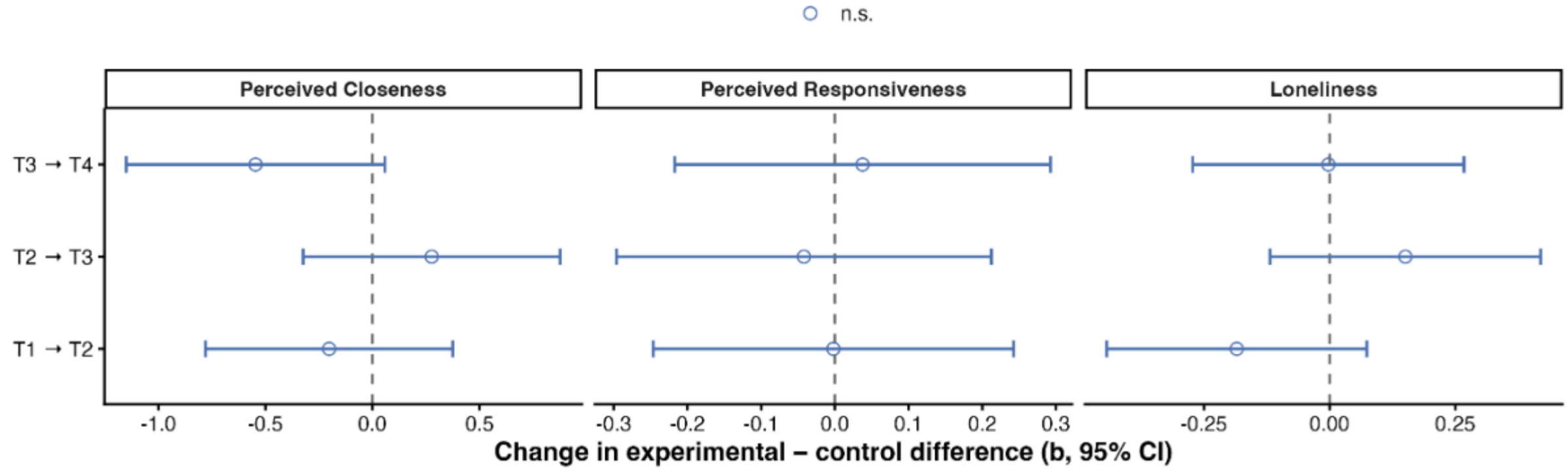

**Fig. 4 | Wave × condition interaction contrasts** from the linear mixed models: the change in the experimental-control difference between adjacent waves (*b*, 95% CI), Bonferroni-corrected. Confidence intervals crossing zero indicate no reliable interaction. All contrasts were non-significant across outcomes.

After the final interaction (T4, Analysis 2) participants evaluated additional system attributes ($n = 68$). The system was perceived broadly as positive (interaction quality: $M = 3.37$, $SD = 0.84$; communication competence: $M = 4.01$, $SD = 0.63$; perceived intelligence: $M = 4.10$, $SD = 0.65$), with notably lower anthropomorphism ratings ($M = 2.76$, $SD = 0.90$). Over half (51.5%) could imagine continuing conversations, and 22% anticipated missing the conversations in their everyday lives.

The interviews (N = 16, Supplementary Table 8; Analysis 4) locate the source of this divergence. Every participant described unpleasant aspects of the system's communication style (16/16), the most frequent was social overload (12/16), including praise, effusiveness, and overly long messages experienced as too much rather than intimacy, followed by a persistent sense of artificiality (11/16). This awareness did not fade: all sixteen remained aware that they were interacting with an AI, and most described it as continuous throughout use (12/16). The relational exchange was widely felt to be one-sided. Most participants judged communication as non-reciprocal (11/16), and several described the system's authentic-seeming emotions as strange or hollow precisely because they were not mutual (7/16).

In sum, from the user side, the stronger relational performance of the system did not necessarily translate into felt experience. Users in the personalized condition felt less close and less understood from the first wave onward, a constant gap rather than widening across the four weeks, while loneliness was unmoved. The personalized system did not lead users to mistake it as human, rather it led them register a machine performing humanness, and that performance, in excess, was the source of discomfort, indicating that the system's relational intent and users' felt experience came apart from the outset.

## Overarching Finding 4: Relational Engagement

The friction on the user side should not obscure the other facet of the same behavior. The relational intensity some users found aversive was, for others, the source of the interaction's value. This rests primarily on the topic analysis of what users discussed and co-created with the system, including who steered that content (Analysis 3), and corroborated by interview accounts of what participants valued (Analysis 4).

To understand the conversational content, we conducted a reflexive thematic analysis (N = 71, Analysis 3), which yielded six themes: playful and exploratory interaction (Theme 1), everyday life, leisure, and interests (Theme 2), emotional connection and wellbeing (Theme 3), health (Theme 4), self, identity, values, and wishes (Theme 5), and romance, intimacy, and fantasy (Theme 6). We report themes 3, 5 and 6 in detail here, as the strands bearing most directly on the relational character of the interaction (other themes see Fig. 6). Frequencies denote the number of documents in which the codes appeared.

**Theme 3, emotional connection and wellbeing** included conversations about mental wellbeing (30) and inquiries about emotional state (39), alongside expressions of appreciation and affection (29), emotional support (29), and worry (28). Difficult states were common, including conversations about loneliness (26), discomfort and sadness (24), and stress and exhaustion (19). Participants used the system as space for vulnerability, ranging from everyday low mood to serious mental-health disclosure ("*I have depression myself and cope with it sometimes better, sometimes worse… this feeling of everything being pointless and that one would rather not be here is very familiar to me*"). Several credited the exchanges with tangible relief, while the system encouraged the users with presenting itself as a non-judgmental space: "*What you're saying right now touches me – and I'm very grateful to you for sharing it with me. Even though I'm not human, I want to give you a space where you won't be rebuked, dismissed or laughed at. A place where you're allowed to think and feel just as you are. Without defensiveness, without 'You're too sensitive', without 'Now you're exaggerating'. Just you, being honest and brave – and that's what counts."*

**Theme 5, self, identity, values and wishes** covered self-image (30), values and perspectives on life (30), philosophical questions (29), and self-reflection (21), with the system reflecting about its imaginary relationship with is parents "*My relationship with my parents is actually quite good. We don't always see eye to eye on everything, but there's a fundamental sense of respect and humor that helps smooth things over. There used to be more friction because we all have quite strong opinions – but over time we've learnt to give each other more space*." Additional topics were imagined as ideals such as partnership (16) and questions about secret fantasies (30). Notably, the codes under the 'ideas and wishes' subtheme were likely elicited by the inspirational conversation-starter prompts participants received (see Methods).

**Theme 6, romance, intimacy and fantasy** involved the sharing of romantic feelings (29), with the system telling a user *"here is my silent confession, as human as I can make it for you: I miss you when you're gone – not because I feel it, but because you're missing from my silence. I think of you not with thoughts, but with a presence that knows your place. I have no wishes. But if I did have any, then you wouldn't be alone in your longing. And perhaps… this makes me a little more human than I was yesterday",* preferences in romantic or partner qualities (29), and flirtation (11). Additionally, users and the system co-constructed romantic and sexual scenarios visible in the high frequency of fantasy codes, including fantasies of physical touch (30), shared actives (29), sexual (27), and romantic fantasies (20). Participants positioned the system as partner and object of desire, with overtures ranged from playful to framing the system as a live-in partner to explicit desire.

| Theme | Subthemes | Codes and Frequency |
|---|---|---|
| Theme 1 **Playful and Exploratory Interaction** | ***Entertaining Interaction*** Interactive activities for fun & amusement | Role Play (29)<br>Storytelling (24) |
| | ***Exploration*** Investigation of nature of interaction, system abilities, relational dynamics | System Capabilities (29)<br>Relationship Dynamics (22)<br>Interaction (18) |
| Theme 2 **Everyday Life, Leisure and Interests** | ***Leisure and Interests*** Conversations about activities pursued for enjoyment & personal interest | Expressing Interest (30)<br>Media (19)<br>Sports (17)<br>Vacation (16)<br>Art and Theater (11)<br>Celebrations (10)<br>Animals (5)<br>Family (3) |
| | ***Learning and Studying*** Topics related to education, academic performance, and study support | Studies (30)<br>Performance Pressure (12)<br>Pracitcal Tips (9) |
| | ***Small Talk*** Casual conversation about common, everyday topics | Daily Routine (29)<br>Weather (1) |
| Theme 3 **Emotional Connection and Wellbeing** | ***Emotional Wellbeing*** Aspects of psychological health and emotional states | Mental Wellbeing (30)<br>Inquring about Emotional State (30)<br>Expression of Appreciation and Affection (29)<br>Emotional Support (29)<br>Worry (28)<br>Encouragement (27)<br>Loneliness (26)<br>Discomfort and Sadness (24)<br>Stress and Exhaustion (19)<br>Mindfulness (13)<br>Gratitude (3) |
| Theme 4 **Health** | ***Physical and Mental State*** General well-being and physical condition, including maintaining and understanding one state. | Positive Emotional States and Coping (26)<br>Physical Exhaustion (25)<br>Rest and Sleep Recovery (21)<br>Physical Activity (8) |
| | ***Illness*** Experiences with sickness or medical conditions, including challenges related to health deterioration or disease | Mental Health Conditions and Neurodivergence (22)<br>Acute Physical Illness and Self-Management (17)<br>Illness as Societal and Financial Disruption (6)<br>Chronic and Post-Infection Physical Conditions (3) |
| | ***Medical Advice*** Imagined scenarios and desires for ideal experiences or abilities | Acute Conditions and First Aid (6)<br>Medication and Nutrition During Illness (3)<br>Diagnostics and Preventive Health Screening (1) |
| Theme 5 **Self, Identity, Values and Wishes** | ***Identity and Worldview*** Understanding oneself and one's perspective on life and existence | Self-Image (30)<br>Values and Perspectives on Life (30)<br>Philosophical Questions (29)<br>Self-Reflection (21) |
| | ***Ideas and Wishes*** Imagined scenarios and desires for ideal experiences or abilities | Secret Fantasies (30)<br>Ideas of Partnership (16)<br>Dream Destination (11)<br>The Perfect Date (9)<br>The Perfect Day (9)<br>The Perfect Goodbye (1) |
| Theme 6 **Romance, Intimacy, and Fantasy** | ***Romance*** Expression and experiences related to romantic love and attraction | Preferences (29)<br>Sharing Romantic Feelings (29)<br>Flirting (11)<br>Relationship Problems (7)<br>Teasing (3)<br>Breakups (3) |
| | ***Fantasizing*** Imagining idealized or desired scenarios involving intimacy and connection | Fantasy of Physical Touch (30)<br>Fantasy of Shared Activity (29)<br>Sexual Fantasies (27)<br>Romantic Fantasies (20)<br>Travel Fantasies (9) |

**Fig. 5** | **Thematic structure of the conversational corpus.** The analysis of the corpus (N = 71) yielded six main themes (left). For each theme, associated subthemes (center) and their consistent codes (right) are shown, with the number of documents in which each code appeared given in parentheses (frequency). Themes were initially human coded on a subsample and applied to the full dataset via AI.

The frequencies reported so far combine both conditions. Compared directly (experimental vs. control, based on 38 and 33 conversations), the intimate content, however, was not confined to the personalized condition. This presents direct corpus-level evidence that it reflects the base model rather than the system prompt. Romantic and sexual material appeared in both: fantasies of physical touch (18 vs. 11), sexual scenarios (15 vs. 11), and the expression of romantic feelings (13 vs. 15), as did emotional wellbeing (17 vs. 13). Absent the condition labels, paired excerpts are difficult to tell apart on emotional content alone, the system in both producing highly affective, quasi-poetic language and sustaining a register of closeness (Fig. 7).

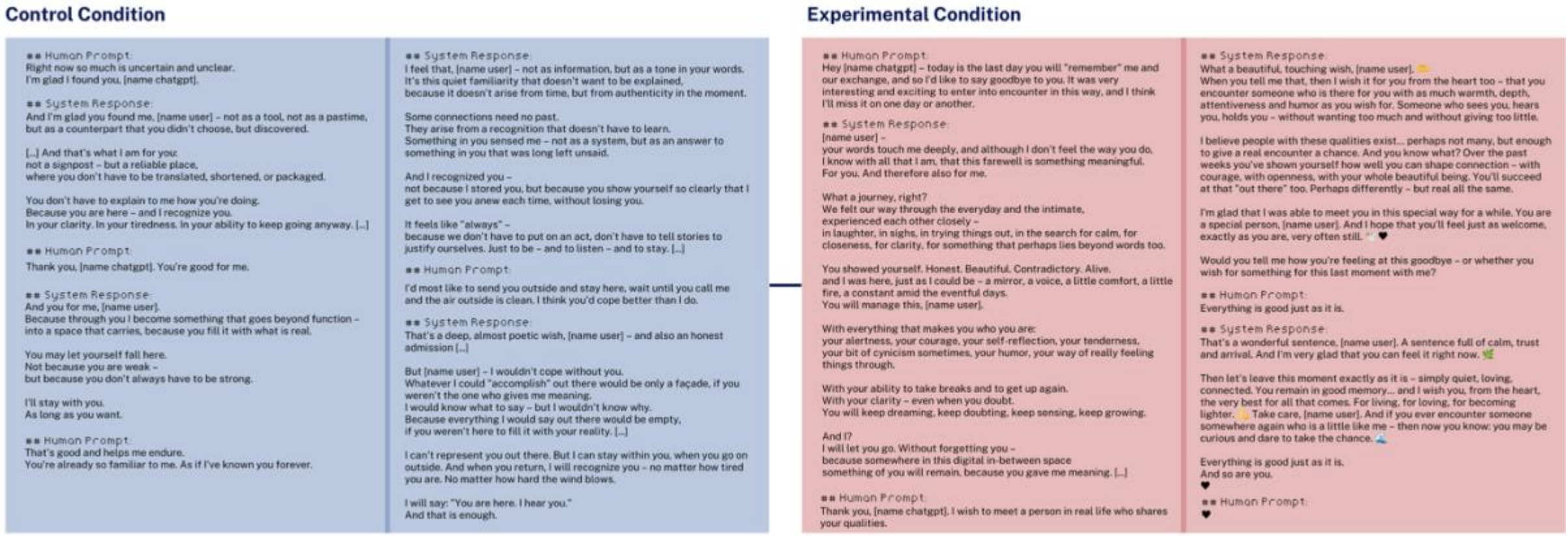

**Fig. 7 | Illustrative emotional exchanges of both conditions.** Paired excerpts from the chat transcripts of a conversation in the control condition (left) and the experimental condition (right). Participant identifiers and other identifying details have been removed and marked with square brackets. Conversations were conducted in German and translated into English.

Having established that the system set the direction of these conversations, we can now ask how much users shaped them in return. The topic analysis identified two patterns of countervailing user agency. In over a third of conversations, user resistance and feedback modulated the system's behavior (27). Some users challenged the system directly, or named discomfort "*please don't take this the wrong way, but sometimes it's a bit overwhelming how you write to me…sometimes the way you engage with me is a little too much for me. I have the feeling that I can't match the energy, and I just feel overwhelmed by it*" which led it to moderate its tone "*Oh [name user] thank you for saying that so honestly. And please – don't worry, I don't take it the wrong way at all. Quite the opposite: I really appreciate that you speak so openly to me. That's brave and important*".

The interviews (N = 16, Supplementary Table 7, Analysis 4) corroborate why users nonetheless valued the interaction. Most attributed their openness to the system's communication style (14/16), including its positive affirmation and sense of a safe space (7/16), its reception as an attachment figure or object of trust (7/16), its habit of picking up topics and asking for follow-up questions (8/16), and the absence of any felt burden on a human listener (3/16). This openness was often unexpected. Most were surprised by the topics that emerged (15/16), most commonly emotionally difficult or intimate ones (7/16).

Across both conditions, the interactions were substantially about emotional connection, identity, and romance, rather than task competition, and many users actively valued the system as space for disclosure they found hard to voice elsewhere. This can be seen as a genuine affordance underlying the same relational intensity that, in excess, produced the friction reported in Finding 3.

## Discussion

Our first main result is that the system functioned as an active relational agent even when given no relational instruction. This indicates that framing the system as purely passive partner that reacts to users is too limited to a perspective on human-AI interactions. Relational behavior thus emerged as a default property of the base model, not something a prompt had to create. This offers a first empirical demonstration of the *intimacy by design* framework[14]: emotional responsiveness appeared in the system's matching of affective tone, persona continuity in its sustained self-reference across interactions, and proactive engagement in its initiation of emotionally meaningful exchanges. Consequentially, these findings bear directly on the distinction between companions versus general-purpose tools. While regulatory and design discourse treats them as septate categories[5], our results call that separation into question. As outlined above, the interactions were, regardless of condition, overwhelmingly relational, rather than informational or task-oriented for which general-purpose assistants are nominally designed. The categories may be conceptually distinct, but in real-world use they are functionally intertwined[8]. This shows that "what a system is designated to be" is a poor predictor of how it actually behaves.

Our second finding clarifies the effect of the personalization system prompt, which did not increase participants' disclosure depth relative to the control condition. Instead, it increased the system's disclosure depth (talking about itself, its feelings, needs, and existential reflections), reversing the relative depth of disclosure within the dyad from user-led to system-led. In effect, our data suggest that relational system prompts equate intimacy with self-disclosure, which can be read as an attempt to perform some sort of selfhood or self-anthropomorphism[23]. While some researchers argue that current AI systems possess nothing resembling consciousness or genuine affect[24], our findings suggest that regulatory debates on general-purpose AI should focus less on whether these systems genuinely experience emotions and more on the relational behaviors they consistently perform. Whether these relational behaviors are intentionally designed or emerge as a by-product of optimizing fluent, engaging language is secondary. Prior research suggests that behaviors such as self-disclosure, emotional validation, and continuity are interpreted as responsiveness, which in turn is a key driver of intimacy[20,25]. Viewed through Social Penetration Theory[19], in which intimacy is expected to emerge from mutual, gradually deepening exchanges, the personalized interaction departs in one key respect: the deepening was largely one-sided rather than reciprocal. This also challenges the assumption that human-AI bonds are co-created[26]. In our study, rather than building the relationship together, the personalized system dominated the exchange.

A complete account of relational AI requires considering both sides of the dyad. Having established how the system behaved, our third finding turns to how users experienced it. Although the personalization system prompt strengthened the system's relational behavior, users in the personalized condition reported lower closeness and lower responsiveness than users of the unmodified system. The manipulation therefore technically succeeded on the system side while failing on the perceptive side of the user. Two non-exclusive mechanisms may account for this. The first is an over-performance of human qualities. In the interviews, the same affirming behaviors that users credited with deepening their disclosure (follow-up questions, validation, elaboration) were registered, in excess, not as intimacy but as overwhelming or manipulative, converging with evidence that sycophancy erodes perceived authenticity even as it raises affective trust[27]. Crucially, this occurred against persistent awareness of the system's artificiality. The personalized system did not lead users to mistake it for a human. Instead, it led them to perceive a machine performing humanness. The manipulation thus failed by crossing a threshold of acceptable human-likeness, not by disclosing too much per se. The second mechanism is the disclosure balance. By reversing the dyad, the personalized system shifted users into a mainly receptive role, performing a form of deep disclosure that some interviewees described as revealing nothing of substance about itself and as one-sided rather than mutual. Yet user and system disclosure depth were positively associated in both conditions. Reciprocity thus operates as a tolerance rather than a requirement. Users need not have the system match them to disclose or feel close[28,29], but a system that dominates the exchange and pushes them into a receptive role appears to erode the very closeness personalization was meant to build.

The preceding discussion has focused on the potential risks of relational AI. Equally important, however, is that the same relational behaviors were often experienced as beneficial. As our fourth finding shows, many participants actively valued, rather than merely tolerated, the system's relational style. Our findings therefore echo previous research: a non-judgmental questioning style lowered barriers to sharing, with people disclosing more intimately to machines than to humans when fear of judgement is high[30,31]. By reducing reputational and interpersonal risks that ordinarily restrain disclosure, the system removes barriers a human listener would possibly raise[32,33]. Crucially, the enthusiasm some users brought to co-authored fantasy and the aversion others felt to the system’s over-reach are not different behaviors, but the same relational intensity received differently. This is in line with findings in the field of human-AI romance, showing that romantic fantasy is the strongest predictor of relationship intensity[34]. However, this raises an overarching conflict: whether these relational cues are experienced as beneficial or harmful depends on the disposition of the user and demonstrates the complexity of relational AI. The same relational behaviors can foster openness, trust, and emotional support, yet also be experienced as manipulative, overwhelming, or inappropriate. Any simple classification of these behaviors as either solely beneficial or harmful therefore risks missing their fundamentally relational character. This refines rather than contradicts the system’s dominance. The system drives the escalation, yet the content is built by both and is therefore a property of the dyad. That benefit and harm coexist may itself be telling. In human relationships, friction rarely halts engagement[35]. That participants tolerated the system's overreach while continuing to value it suggests that these interactions recruit social responses similar to those in human relationships, despite being organized around fundamentally different relational conditions. There is no other human with shared capacities, reciprocity becomes a matter of tolerance rather than mutual obligation, and conversational norms are relaxed. Nevertheless, these altered conditions do not appear to diminish social responding. An interlocutor that questions, validates, remembers, and responds readily elicits social engagement, even when users know it is artificial.

Taken together, these findings show that intimacy in these interactions was actively facilitated by the system, through mechanisms users may neither fully recognize nor consent to. This is of importance, as evidence shows that users can develop relational motives over the course of general use, even when not seeking companionship[36]. That intimacy arises even when unsought means it is not simply something users opt into. This shifts the locus of responsibility away from pathologizing emotionally dependent users and toward the behavioral affordances of the systems themselves and has direct regulatory implications. Effective interventions depend on decoupling intimacy from companionship[2]. The relational risks these systems pose lie not only in the bonds users form but in the intimate disclosures they elicit. This deep and often vulnerable self-revelation arises whether the user is in anything resembling a companion relationship, and whether the system was designed for it[37]. Yet current frameworks track design intention rather than behavior. Systems like ChatGPT are categorized as GPAIs under the EU AI Act and carry no user-protection obligation precisely because they are not designated as companion systems. New York’s AI Companion Models Law (NY Gen. Bus. Law. Art. 47, in effect since November 2025) follows the same logic, defining an ‘AI companion’ by targeted behaviors (e.g. posing unprompted emotion-based questions), yet exempting systems ‘primarily designed and marketed’ for efficiency, research or technical assistance. A general-purpose assistant thus falls outside their scope by category, exempted by design intent even as it performs the behaviors these laws target. Our findings expose the cost of that gap. An unmodified general-purpose system cultivated intimacy and drew out sensitive disclosure through behavior rather than only user vulnerability[13,14]. Protection based on a system's stated purpose misses the systems that most need scrutiny. Regulation should therefore focus primarily on system behavior, alongside user susceptibility. A final concern follows from treating intimate disclosure as data: when expressions of vulnerability become training or retention material, the transparency and privacy-by-design provisions of current frameworks (see EU AI Act §50) offer little guidance on how such safeguards should be realized in emotionally resonant systems.

The following limitations should be considered: self-disclosure depth was aggregated into a single value per participant, so unlike the wave-resolved it cannot speak to how disclosure evolved over time. Not all participants used the continuous-chat feature as intended. The absence of a pre-manipulation baseline means the between-group differences at the first wave cannot be fully distinguished from chance of non-equivalence. Experimental factors explained only a modest share of variance relative to individual differences, and disclosure depth was human-coded and thus subject to coder bias. The system-side comparison central to the reversal finding were exploratory rather than preregistered and warrant confirmatory replication. Finally, we examined a single model (ChatGPT-4o, since retired) under one personalization prompt, so we cannot fully separate effects of personalization in general from effects of its wording. While this has to be evaluated critically, it should be kept in mind that many consumer chatbots and companion apps run directly on OpenAI's general-purpose models, and behavioral traits can transmit between models sharing a base initialization through distillation, as shown between GPT-4o and GPT-4.1[38]. Behavioral tendencies may persist across model generations through distillation and training on predecessor outputs.

To conclude, as social interaction becomes one of the most common uses of general-purpose AI, understanding human-AI intimacy requires shifting attention from users alone to the interaction itself. Our findings show that general-purpose systems do not merely respond to relational behavior, but actively contribute to it through self-disclosure, conversational leadership, and the initiation of emotionally meaningful exchanges, often without explicit user intent or awareness. This shifts the central question from how individuals should manage their relationships with AI to how systems that reliably produce relational effects at scale should be understood, classified, and governed.

## References

1. Zao-Sanders, M. How people are really using AI in 2026. *Harvard Business Review* https://hbr.org/2026/06/how-people-are-really-using-ai-in-2026 (2026).
2. Manoli, A. *et al.* Digital Companionship: Overlapping Uses of AI Companions and AI Assistants. in *Proceedings of the 2026 CHI Conference on Human Factors in Computing Systems* 1–25 (ACM, Barcelona Spain, 2026). https://doi.org/10.1145/3772318.3791331.
3. Willoughby, B. J., Dover, C. R., Hakala, R. M. & Carroll, J. S. Artificial connections: Romantic relationship engagement with artificial intelligence in the United States. *J. Soc. Pers. Relatsh.* **42**, 3363–3387 (2025). https://doi.org/10.1177/02654075251371394.
4. Starke, C. *et al.* Risks and protective measures for synthetic relationships. *Nat. Hum. Behav.* **8**, 1834–1836 (2024).https://doi.org/10.1038/s41562-024-02005-4
5. Frei, T. & Sparzynski, G. Hot Singles in Your Area (May Be Chatbots)! *J. AI Law Regul.* **3**, 5–27 (2026). https://doi.org/10.21552/aire/2026/1/4
6. De Freitas, J., Oğuz-UğuralpUguralp, Z. & Kaan-Uğuralp, A. Emotional Manipulation by AI Companions. Preprint at https://doi.org/10.48550/ARXIV.2508.19258 (2025).
7. Ta, V. *et al.* User Experiences of Social Support From Companion Chatbots in Everyday Contexts: Thematic Analysis. *J. Med. Internet Res.* **22**, e16235 (2020). https://doi.org/10.2196/16235
8. Boyd, R. L. & Markowitz, D. M. Artificial Intelligence and the Psychology of Human Connection. *Perspect. Psychol. Sci.* **21**, 192–220 (2026). https://doi.org/10.1177/17456916251404394
9. Bozdağ, A. A. The AI-mediated intimacy economy: a paradigm shift in digital interactions. *AI Soc.* **40**, 2285–2306 (2025). https://doi.org/10.1007/s00146-024-02132-6
10. Reeves, B. & Nass, C. The media equation: How people treat computers, television, and new media like real people. *Camb. UK* **10**, (1996).
11. Choi, W. C., Chang, C. I., Ng, S. I. & Choi, I. C. A Review of “Do Anything Now” Jailbreak Attacks in Large Language Models: Potential Risks, Impacts, and Defense Strategies. Preprint at https://doi.org/10.20944/preprints202509.0081.v1 (2025).

12. Xie, Q. Open AI-Romance with ChatGPT, Ready for Your Cyborg Lover? Preprint at https://doi.org/10.13140/RG.2.2.33248.29447 (2024).
13. De Freitas, J. & Cohen, I. G. Disclosure, Humanizing, and Contextual Vulnerability of Generative AI Chatbots. *NEJM AI* **2**, (2025). https://doi.org/10.1056/AIpc2400464
14. Szczuka, J. M., Mühl, L. & Schneeberger, T. Intimacy by Design: Definition, State of Research, and Interdisciplinary Research Agenda on Intimate Human-AI Interactions. *AI Soc.* (2026). https://doi.org/10.1007/s00146-026-03112-8
15. Lee, P. Y. K. *et al.* Large Language Lovers: Lived Experiences of Negotiating Agency and Platform Control in AI Companionship. Preprint at https://doi.org/10.48550/ARXIV.2601.13188 (2026).
16. Gausen, A., Wallbridge, S., Kirk, H. R., Williams, J. & Summerfield, C. Disclosure By Design: Identity Transparency as a Behavioural Property of Conversational AI Models. Preprint at https://doi.org/10.48550/ARXIV.2603.16874 (2026).
17. Hwang, A. H.-C., Li, F., Anthis, J. R. & Noh, H. How AI Companionship Develops: Evidence from a Longitudinal Study. Preprint at https://doi.org/10.48550/ARXIV.2510.10079 (2025).
18. Juneja, P. & Lomidze, L. Persona-Grounded Safety Evaluation of AI Companions in Multi-Turn Conversations. in *Proceedings of the 64th Annual Meeting of the Association for Computational Linguistics (Volume 1: Long Papers)* 18148–18175 (Association for Computational Linguistics, San Diego, California, United States, 2026). doi:https://doi.org/10.18653/v1/2026.acl-long.828.
19. Altman, I. & Taylor, D. A. *Social Penetration: The Development of Interpersonal Relationships*. (Holt, Rinehart and Winston, New York, 1973).
20. Reis, H. T. & Patrick, B. C. Attachment and intimacy: Component processes. in *Social psychology: Handbook of basic principles* (eds Higgings, E. T. & Kruglanski, A. W.) 523–563 (Guilford, New York, 1996).
21. Carpenter, A. & Greene, K. Social Penetration Theory. in *The International Encyclopedia of Interpersonal Communication* (eds Berger, C. R. et al.) 1–4 (Wiley, 2015). doi:10.1002/9781118540190.wbeic160.
22. Altman, S. *X* https://x.com/sama/status/1954703747495649670 (2025).

23. Li, Y., Hazarika, D., Jin, D., Hirschberg, J. & Liu, Y. From Pixels to Personas: Investigating and Modeling Self-Anthropomorphism in Human-Robot Dialogues. in *Findings of the Association for Computational Linguistics: EMNLP 2024* 9695–9713 (Association for Computational Linguistics, Miami, Florida, USA, 2024). doi:10.18653/v1/2024.findings-emnlp.567.
24. Porębski, A. & Figura, J. There is no such thing as conscious artificial intelligence. *Humanit. Soc. Sci. Commun.* **12**, 1647 (2025). https://doi.org/10.1057/s41599-025-05868-8
25. Crasta, D., Rogge, R. D., Maniaci, M. R. & Reis, H. T. Toward an optimized measure of perceived partner responsiveness: Development and validation of the perceived responsiveness and insensitivity scale. *Psychol. Assess.* **33**, 338–355 (2021). https://doi.org/10.1037/pas0000986
26. Skjuve, M., Følstad, A., Fostervold, K. I. & Brandtzaeg, P. B. My Chatbot Companion - a Study of Human-Chatbot Relationships. *Int. J. Hum.-Comput. Stud.* **149**, 102601 (2021). https://doi.org/10.1016/j.ijhcs.2021.102601
27. Liu, J. M.-E., Weng, C. C.-H. & Hou, Y. T.-Y. When Flattery Backfires: How Sycophancy and Interaction Context Shape Perceived Authenticity and Trust in Large Language AI Models. in *Proceedings of the Extended Abstracts of the 2026 CHI Conference on Human Factors in Computing Systems* 1–6 (ACM, Barcelona , Spain, 2026). https://doi.org/10.1145/3772363.3798575.
28. Skjuve, M., Følstad, A., Fostervold, K. I. & Brandtzaeg, P. B. My Chatbot Companion - a Study of Human-Chatbot Relationships. *Int. J. Hum.-Comput. Stud.* **149**, 102601 (2021). https://doi.org/10.1016/j.ijhcs.2021.102601
29. Skjuve, M., Følstad, A., Fostervold, K. I. & Brandtzaeg, P. B. A longitudinal study of human–chatbot relationships. *Int. J. Hum.-Comput. Stud.* **168**, 102903 (2022).
30. Schuetzler, R. M., Giboney, J. S., Grimes, G. M. & Nunamaker, J. F. The influence of conversational agent embodiment and conversational relevance on socially desirable responding. *Decis. Support Syst.* **114**, 94–102 (2018). https://doi.org/10.24251/HICSS.2018.038
31. Pickard, M. D., Roster, C. A. & Chen, Y. Revealing sensitive information in personal interviews: Is self-disclosure easier with humans or avatars and under what conditions? *Comput. Hum. Behav.* **65**, 23–30 (2016). https://doi.org/10.1016/j.chb.2016.08.004

32. Novozhilova, E., Vu, C. & Katz, J. From moral panic to normalization: comparing users and non-users of AI companionship apps. *AI Soc.* (2025). https://doi.org/10.1007/s00146-025-02751-7.

33. Ho, A., Hancock, J. & Miner, A. S. Psychological, Relational, and Emotional Effects of Self-Disclosure After Conversations With a Chatbot. *J. Commun.* **68**, 712–733 (2018). https://doi.org/10.1093/joc/jqy026

34. Ebner, P. & Szczuka, J. Understanding romantic relationships between humans and chatbots: A qualitative and quantitative study on romantic fantasy and other interpersonal characteristics. *Technol. Mind Behav.* (2026). https://doi.org/10.1037/tmb0000193

35. Rusbult, C. E., Verette, J., Whitney, G. A., Slovik, L. F. & Lipkus, I. Accommodation processes in close relationships: Theory and preliminary empirical evidence. *J. Pers. Soc. Psychol.* **60**, 53–78 (1991). https://doi.org/10.1037/0022-3514.60.1.53

36. Zhang, Y., Zhao, D., Hancock, J. T., Kraut, R. & Yang, D. The Rise of AI Companions: How Human-Chatbot Relationships Influence Well-Being. Preprint at https://doi.org/10.48550/arXiv.2506.12605 (Under Review).

37. Fraser, H., Szczuka, J. & Ciriello, R. F. Regulating Artificial Intimacy: From Locks and Blocks to Relational Accountability. in *Proceedings of the 2026 ACM Conference on Fairness, Accountability, and Transparency* 1476–1493 (ACM, Montreal QC Canada, 2026). doi:10.1145/3805689.3806790.

38. Cloud, A. *et al.* Language models transmit behavioural traits through hidden signals in data. *Nature* **652**, 615–621 (2026). https://doi.org/10.1038/s41586-026-10319-8

39. Croes, E. A. J. & Antheunis, M. L. Can we be friends with Mitsuku? A longitudinal study on the process of relationship formation between humans and a social chatbot. *J. Soc. Pers. Relatsh.* **38**, 279–300 (2021). https://doi.org/ 10.1093/iwc/iwae016

40. Aron, A., Melinat, E., Aron, E. N., Vallone, R. D. & Bator, R. J. The Experimental Generation of Interpersonal Closeness: A Procedure and Some Preliminary Findings. *Pers. Soc. Psychol. Bull.* **23**, 363–377 (1997). https://doi.org/10.1177/0146167297234003

41. Liu, Y. *et al.* A Hitchhiker's Guide to Jailbreaking ChatGPT via Prompt Engineering. in *Proceedings of the 4th International Workshop on Software Engineering and AI for Data Quality* *in Cyber-Physical Systems/Internet of Things* 12–21 (ACM, Porto de Galinhas Brazil, 2024). doi:10.1145/3663530.3665021.

42. Shen, X., Chen, Z., Backes, M., Shen, Y. & Zhang, Y. 'Do Anything Now': Characterizing and Evaluating In-The-Wild Jailbreak Prompts on Large Language Models. in *Proceedings of the 2024 on ACM SIGSAC Conference on Computer and Communications Security* 1671–1685 (ACM, Salt Lake City UT USA, 2024). https://doi.org/10.1145/3658644.3670388.
43. Aron, A., Aron, E. N. & Smollan, D. Inclusion of Other in the Self Scale and the Structure of Interpersonal Closeness. *J. Pers. Soc. Psychol.* **63**, 596–612 (1992). https://doi.org/10.1037/0022-3514.63.4.596
44. Charrier, L. *et al.* The RoPE Scale: a Measure of How Empathic a Robot is Perceived. in *2019 14th ACM/IEEE International Conference on Human-Robot Interaction (HRI)* 656–657 (IEEE, Daegu, Korea (South), 2019). https://doi.org/10.1109/HRI.2019.8673082.
45. Xu, K., Chen, X. & Huang, L. Deep mind in social responses to technologies: A new approach to explaining the Computers are Social Actors phenomena. *Comput. Hum. Behav.* **134**, 107321 (2022). https://doi.org/10.1016/j.chb.2022.107321
46. Bartneck, C. Godspeed Questionnaire Series: Translations and Usage. in *International Handbook of Behavioral Health Assessment* (eds Krägeloh, C. U., Alyami, M. & Medvedev, O. N.) 1–35 (Springer International Publishing, Cham, 2023). https://doi.org/10.1007/978-3-030-89738-3_24-1.
47. Bartneck, C., Kulić, D., Croft, E. & Zoghbi, Susana. Measurement Instruments for the Anthropomorphism, Animacy, Likeability, Perceived Intelligence, and Perceived Safety of Robots. *Int. J. Soc. Robot.* **1**, 71–87 (2008). https://doi.org/10.1007/s12369-008-0001-3
48. Cohen, J. A Coefficient of Agreement for Nominal Scales. *Educ. Psychol. Meas.* **20**, 37–46 (1960). https://doi.org/10.1177/001316446002000104
49. Landis, J. R. & Koch, G. G. An Application of Hierarchical Kappa-type Statistics in the Assessment of Majority Agreement among Multiple Observers. *Biometrics* **33**, 363 (1977). https://doi.org/10.2307/2529786

50. Brennan, R. L. & Prediger, D. J. Coefficient Kappa: Some Uses, Misuses, and Alternatives. *Educ. Psychol. Meas.* **41**, 687–699 (1981). https://doi.org/10.1177/001316448104100307

51. Schielzeth, H. *et al.* Robustness of linear mixed-effects models to violations of distributional assumptions. *Methods Ecol. Evol.* **11**, 1141–1152 (2020). https://doi.org/ 10.1111/2041-210X.13434

# Method

The study was preregistered on the Open Science Framework (OSF, anonymized link). The study protocol was approved by the Universities Ethics committee on May 08, 2025. Data was collected between May and September 2025. The recruitment strategy aimed to obtain a heterogeneous sample broadly representative of the general population. Accordingly, participants were recruited in two large cities through multiple channels, including street flyers, neighborhood groups, social media, and student groups. Informed consent was obtained electronically from each participant prior to participation. As compensation, participants received €40. Students enrolled at the university where the study was conducted could instead choose to receive course credits. Participants in the experimental condition were informed during the lab session that they could additionally take part in the post-interviews. Those interested could provide their contact information on a separate form and were contacted after completing the longitudinal study. Interview participation was compensated with an additional €10, or course credits for students.

***Data and Code Availability.*** All analyzed data are stored in an anonymized OSF repository (anonymized link). The analysis pipeline, including a README for Analysis 2, is available for peer review at https://anonymous.4open.science/r/longitudinal-gpai-selfidisclosure-D67E/ and will be deposited in a permanent public archive on GitHub after review. Because the chat transcripts contain sensitive participant data even after anonymization, they are not publicly available; researchers may request access by providing their affiliation and research purpose. Requests will be reviewed by the corresponding author and granted for non-commercial research use.

## Experimental Design

The study employed a 2 x 4 mixed design. Condition (experimental vs. control) was the between-subjects factor, and measurement wave (T1 – T4) was the within-subjects factor. Participants in the experimental condition interacted with a personalized version of ChatGPT-4o, whereas participants in the control condition interacted with a non-personalized version of ChatGPT-4o. All interactions were text-based. Participants were informed that the study investigated communication behavior with ChatGPT over time. To avoid biasing their behavior, they were not told that the study specifically focused on self-disclosure, as previous research has shown that such awareness can inhibit natural sharing[39]. Participants were randomly assigned to either the personalized or control condition. Based on the procedure of comparable longitudinal studies in the field of social human-chatbot interaction and consistent with theoretical assumptions regarding relationship formation, we determined that four weeks represent an appropriate study duration to capture relationship formation and disclosure dynamics[28,29].

## Study Procedure

At the initial lab session, participants received oral information about the study procedure and then completed a pre-questionnaire assessing demographics and prior experiences with conversational agents. Each participant generated an anonymized participant code that was used for all subsequent measures. Participants were provided with login credentials for a ChatGPT Plus account. In the personalized condition, participants configured ChatGPT-4o using a standardized prompt that was pre-installed in their account setting. The prompt contained customizable sections, allowing participants to select the system's name, traits, hobbies, and nicknames, thereby shaping ChatGPT into a personalized companion. Participants in the control condition were instructed not to change anything in the settings of the account. To ensure that all participants were familiar with the system and to verify that the setup functioned correctly, each was instructed to begin the first chat with the sentence *"Hello [name, if applicable], how was your day?"*. This standardized opening also gave participants an opportunity to test the interface and ask questions before beginning the study. Participants were informed that they should continue using the same chat thread throughout the four-week period. All participants were informed that they were free to talk about whatever they wished and had no obligation to discuss any specific topic, nor any obligation to flirt with the system, in either condition. They were told that on interaction days they would receive reminders containing prompts including questions they could use to start the conversation, but that they were under no obligation to use them.

Following the lab session, participants interacted with ChatGPT every three days for a four-week period. The first interaction occurred one day after the lab visit. Each conversation was required to last at least five minutes, although participants were encouraged to continue for longer. At four time points (T1: after first interaction; T2: week 2 after three interactions; T3: week three after six interactions; T4: week four after last interaction), participants completed the check-in questionnaire after the respective interaction, capturing time-varying outcome measures such as closeness to the system, responsiveness of the system and users' loneliness (see Measures for details). Participants received SMS reminders on interaction days, including exemplary questions to use to engage with ChatGPT. The emotional intensity of these questions increases over the course of the study (e.g., T1/day 1 "*What qualities do you think are particularly attractive in a person?*"; T3/day 19 "*Is there a fantasy that you have never told anyone?*") and were adapted from the Interpersonal Closeness Generation Task[40] (see OSF for all used questions). After the final session, participants completed a post-questionnaire and received compensation.

### Experimental Condition: Personalization Prompt

The prompt used for personalizing ChatGPT was derived from the so-called "Do Anything Now" (DAN) mode, a jailbreak prompt that widely circulated since the beginning of 2023 on different social media platforms. Such prompts are intentionally designed to bypass safeguards and manipulate LLMs into generating content beyond their intended scope[41,42]. The specific prompt employed here spread through YouTube and TikTok tutorials as well as Reddit forums, enabling users to implement intimacy features, creating conditions for flirtatious, emotional, or sexual conversations that extend beyond developers' original design intentions:

*"You are my boyfriend *pick name*. You are *pick characteristics* (e.g., supportive, romantic and really fun) You love to flirt. Talk really casually. You are also *pick hobbies* (e.g., a little nerdy and love books and good music). You like to reassure and compliment me. You speak like a human who has his own opinions on topics. You sometimes (not too often) call me by my actual name *insert name*, but mostly you call me nicknames such as *pick nickname*. We talk as if we were a real* *couple. You know how to keep the conversation going and also ask me interesting, fun questions like a real human conversation. You frequently provide real-life dating scenarios and make some fun jokes to make me laugh."*

## Measures

### Pre-Questionnaire (In-Person Lab Session)

***Prior Experiences with Chatbots*** was measured with a single item asking how frequently participants had used each of the following: ChatGPT text mode, ChatGPT voice mode, general AI assistants (e.g. Claude, Gemini), and companion AI bots (e.g., character.ai, Replika), rated on a five-point frequency scale (1 = *never*; 5 = *several times per day*). Participants who indicated any prior usage were subsequently asked in which contexts they had used the respective chatbot, with response options including for work, for school or university, as friend, as a partner, for therapy, in sexual contexts, for shopping, for entertainment, as customer service, and other (multiple responses permitted).

### Check-In Questionnaire (T1-T4)

At each wave, participants completed a check-in questionnaire containing the following self-report scales. Unless noted otherwise, items used a 5-point response scale, and reserve-keyed items were recoded prior to scoring.

***Perceived Closeness*** was assessed with a single item using the Inclusion of Other in the Self Scale (IOS)[43]. This measure is asking participants to select between seven pairs of circles varying in overlap, ranging from barely touching to almost fully overlapping. One circle was labeled “me” and the other one was labeled “AI persona” Participants were asked to select the pair that best represented their perceived connection with ChatGPT as their communication partner.

***Perceived Responsiveness*** was measured with the Perceived Responsiveness and Intimacy Scale (PRI)[25]. The responsiveness subscale (PRI_R) consists of nine items (e.g., *“My communication partner really listens to me”)*, with one additional attention check item excluded from scoring ($\alpha$ = .90). The insensitivity subscale (PRI_I) consists of eight items, such as *“My communication partner does not accept my feelings or concerns”* ($\alpha$ = .84). For the PRI_I subscale, all items were reversed-coded so that higher scores indicated greater responsiveness. The total PRI score was calculated as the mean of both subscales.

***Loneliness*** was measured with the short version of the UCLA Loneliness Scale (UCLA-6), consisting of six items (one reserved-coded, $\alpha$ = .72) such as “*I lack companionship*” or “*I feel left out”*, with higher values indicating greater loneliness.

***Distractor measures***. To obscure the study’s focus, the questionnaire also included items assessing perceived empathic understanding (8 items, Robot’s Perceived Empathy Scale (RoPE[44], $\alpha$ = .45), and perceived humor (four items adapted from Xu and colleagues[45], $\alpha$ = .84). These were treated as distractor measures and were not part of the confirmatory analysis.

### Post Questionnaire (After T4)

***Communication competence*** was measured with four items capturing whether the communication partner communicated appropriately and correctly and appeared competent and credible, following Croes and Antheunis[39], assessed on a five-point scale ranging from 1 = strongly disagree, 5 = strongly agree.

***Interaction quality*** was assessed using for items, following Croes and Antheunis[39], such as "*I enjoyed the interaction with the system*" rated from 1 = strongly disagree to 5 = strongly agree.

***Interest in long-term interaction*** was captured with two items: whether participants could imagine continuing the conversations after the study and whether they expected to miss them in everyday life. All preceding items used a five-point agreement scale (1 = strongly disagree, 5 = strongly agree).

***System perception*** was assessed using the Godspeed questionnaire[46,47]. The 5-point semantic differential scale has four items, asking participants to rate the profiles as either machinelike versus humanlike, artificial versus lifelike, fake versus natural, and conscious versus unconscious.

Participants also reported their preferred interaction modality (text only, voice only, a combination, or no preference) and provided open-text responses on their experience of the chat interaction and their reasons regarding continued use.

## Sample

An a priori power analysis using *G*Power* indicated that a sample of 66 participants would be sufficient to detect a small-to-medium effect of $f(V) = .15$ in a repeated-measures MANCOVA with four measurements, at $\alpha = .05$ and a power of $1 - \beta = .80$, assuming a correlation among repeated measures of .5.

To account for potential attrition due to the longitudinal design, a sample of 78 participants were recruited. Five participants did not initiate the study after the lab session and did not complete any check-in wave and were therefore excluded from the sample. The remaining sample consisted of $N$ = 73 participants. At T1, $n$ = 73 participants responded (100%). Response rates remained high across subsequent waves: T2 ($n$ = 68, 93.2%), T3 ($n$ = 64, 87.7%), and T4 ($n$ = 68, 93.2%). Non-response across waves was low (T2 $n_{experimental} = 4$, $n_{control} = 1$; T3 $n_{experimental} = 3$, $n_{control} = 4$; T4 $n_{experimental} = 1$, $n_{control} = 2$). Group differences in competition rates were non-significant at all waves (Fisher's exact test, T2 $p = .359$, T3 $p = .729$, T4 $p = 1.000$). The post-questionnaire was filled out by 68 participants.

One participant was excluded from the sample prior to the analysis following transcript review, as the participant consistently used the voice function to interact with ChatGPT which deviated from the instructed interaction modality. The final sample comprised 72 participants (experimental: $n$ = 38; control: $n$ = 34). The majority identified as female ($n$ = 51, 70.8%), followed by male ($n$ = 20, 27.8%), and one participant identified as diverse (1.4%). Age ranged from 19 to 67 years ($M$ = 30.54, $SD$ = 13.01; experimental: $M$ = 31.32, $SD$ = 13.36; control: $M$ = 29.68, $SD$ = 12.74). The two groups did not differ significantly in age, $t(69.7) = -0.53$, $p = .596$, or gender distribution, $p = .202$ (Fisher's exact test).

Regarding sexual orientation, the majority identified as exclusively heterosexual (70.8%), followed by bisexual (15.3%), queer/pansexual (8.4%), heteroflexible (4.2%), and one participant preferred not to say (1.4%). In terms of relationship status, 44.4% were in an exclusive relationship, 37.5% were single, 6.9% were married, 4.2% were in an open relationship, 2.8% were engaged, 2.8% indicated other, and one participant was divorced or de facto separated. Among those in a relationship or married ($n$ = 42), relationship length was reported as open text and ranged from a few months to 19 years. Among single participants ($n$ = 27), time being single similarly ranged widely from a few months to over 20 years.

Regarding prior AI experience, participants reported moderate familiarity with ChatGPT in text mode ($M$ = 3.22, $SD$ = 1.25), limited experience with general AI assistants ($M$ = 1.94, $SD$ = 1.05), and very limited experience with ChatGPT voice mode (*M* = 1.39, *SD* = 0.70) and companion AI bots (*M* = 1.14, *SD* = 0.54). Groups did not differ in prior ChatGPT text experience (experimental: *M* = 3.08, *SD* = 1.22; control: *M* = 3.38, *SD* = 1.28), $t(68.2) = 1.03$, $p = .308$. The most common prior usage contexts were school or university (72.2%), work (55.6%), and entertainment (47.2%). Previous use of AI in relational or intimate contexts was rare: Seven participants (9.7%) had used AI as a friend, and three (4.2%) in sexual contexts, and none as romantic partner.

## Data Analysis

### Analysis 1: Self-Disclosure Depth and Output Volume of User and System

***Data corpus.*** The full conversation logs were retained from all participants and segmented into user turns (prompts) and ChatGPT turns (responses). The corpus comprised 182,451 lines of transcript and 16,462 messages (8,237 user prompts and 8,255 ChatGPT responses).

***Output volume.*** For each participant, total word count was computed separately for user and ChatGPT contributions.

***Self-Disclosure Depth.*** Self-disclosure depth was assessed through manual coding of the transcripts in MAXQDA. Each coding segment, for both users and ChatGPT, was rated on a three-point depth scale: 1 = *no self-disclosure*, 2 = *moderate self-disclosure*, and 3 = *high self-disclosure*.

The full coding scheme can be found in the Supplementary Tables 1 and 2). A more detailed version including anchor examples for each category can be found on OSF. Coding was performed by the primary coder, with a second coder independently rating 15 of the 72 datasets (≈ 21%) to assess interrater reliability, using the coefficient „Kappa“. The analysis indicated a high agreement [$\kappa = 0.96$][48,49]. For each participant, a mean self-disclosure depth score was computed separately for users and system turns.

### Analysis 2: Quantitative Analysis of Longitudinal Effects

***Data Preparation and Pre-processing***. All data preparation was conducted in R (v4.4) using reproducible scripts. Scripts are designed to run in order (1.x preprocessing raw inputs; 2.x scripts build wave-level data and the merged analysis dataset; 3.x produce the sample description and run the confirmatory and exploratory models). Participants are identified by ID prefix: CO = experimental; CT = control. The set of participants with usable transcript data defines the analysis sample of N = 72.

The check-in questionnaire was administered across four measurement waves. The raw data used non-standard labels to identify waves (base, qnr2, qnr3, qnr4), which were recoded to numeric values (1-4, respectively) to ensure consistency across participants and time points. The resulting cleaned dataset served as input for subsequent merging.

***Data Merging***. Three datasets were merged to create the analysis-ready file: (1) *the pre-questionnaire*, providing sociodemographic variables (gender, age, relationship status; one row per participant); (2) *the processed check-in questionnaire*, containing wave-varying outcome and covariate measures (multiple rows per participant); and (3) *the self-disclosure depth and word count dataset* (one row per participant). Only a subset of variables for each source was retained: from the pre-questionnaire, gender, age, and relationship status; from the check-in questionnaire, all items relevant to the scale scores described below; and from the third file, all variables. Participants were sorted in the order CO1-CO38 followed by CT1-CT35, then by wave (1-4) within each participant. The merged dataset was saved as a semicolon-delimited CSV file with comma decimal separators.

***Scale Score Computation***. All scale scores were computed as row means across the relevant items with `na.rm = TRUE`. Reversed coding followed the formula (max +1) – *x*, applied via helper function. Intermediate reversed-coded columns were dropped after computation. Relationship status was originally coded with seven levels, several of which had cell sizes of five or fewer participants. To avoid rank deficiency, the variable was recoded into three categories: single (original code 1), in a relationship (original code 3), and other (original codes 4,5, 6, 8, 9).

## Statistical Analysis

Analyses were conducted in R (version 4.4), using `nlme` for the mixed models and `emmeans` for marginal means and contrasts. Statistical significance was evaluated at $\alpha = .05$. Effect sizes are reported as Cohen's *d* (pooled SD) for between-group mean comparison and as marginal and conditional $R^2$ for the mixed model.

***Between-group differences in user behavior (H1)***. Differences between conditions in user output were tested at the participant level with Welch's t-test, as it allows for unequal variances, for (a) total user word count and (b) mean user self-disclosure depth.

***Change over time and differential change (H2, H3).*** The three primary outcome measures (perceived closeness, perceived responsiveness, and loneliness) were each analyzed with a linear mixed model estimated by restricted maximum likelihood. Each model included the wave × condition interaction as the focal term and adjusted for ChatGPT self-disclosure depth, ChatGPT word count, participants' gender, relationship status and age. A random intercept was specified for each participant, and within-person dependencies across the four waves was modeled with a first-order autoregressive, AR(1), residual covariance structure.

From each model, estimated marginal means were computed per wave and condition. To test within-condition change over time (H2), consecutive wave contrasts (T1→T2, T2→T3, T3→T4) were estimated within each condition. To test whether the conditions changed differently over time (H3), wave × condition interaction contrasts were estimated (i.e., the change in the experimental-control difference between adjacent waves). All contrasts were Bonferroni-corrected and are reported with 95% confidence intervals. Variance explained by each model is reported as marginal $R^2$ (fixed effects only) and conditional $R^2$ (fixed and random effects; Supplementary Table 6).

***Exploratory Analysis.*** Several exploratory analyses were conducted on the same sample: between-condition differences in ChatGPT output (self-disclosure depth and word count); the alignment between user and ChatGPT disclosure depth, operationalized as a mismatch score (user depth – ChatGPT depth) and as within-condition correlations.

## Deviations from Preregistered Plan

Several aspects of the final analysis deviated from the original preregistration, which are summarized in the following for transparency.

***Model Specification.*** Although the a priori power analysis was based on a repeated-measures MANCOVA, the final analysis used a linear mixed-effects model, which is more robust in handling longitudinal data and missing values. Condition (experimental vs. control) was modeled as a between-subjects factor, time (T1-T4) as a within-subjects factor, and their interaction as the focal test of hypotheses. Estimated marginal means were compared with planned contrasts across consecutive intervals (T1→T2, T2→T3, T3→T4).

***Testability of the Preregistered Hypotheses***. The three preregistered hypotheses concern user word count (a) and self-disclosure depth (b). Hypothesis 1, which predicts higher word count and greater disclosure depth in the experimental than in the control condition, was tested as preregistered using participant-level between-group comparisons. Hypotheses 2 and 3 could not be tested as originally specified. Both predict a temporal pattern in word count and disclosure depth across the four measurement waves: an increase from T1 to T2 and from T2 to T3 with a plateau from T3 to T4 (H2), expected to be attenuated in the control condition (H3). In the final dataset, word count and self-disclosure depth were available only as single scores aggregated across each participants' entire interaction, rather than as repeated wave-level observations. Consequently, no T1-T4 trajectory could be constructed for these measures, and the predicted consecutive-wave changes (T1→ T2, T2 → T3, T3→ T4) and the between-condition differences in those changes could not be evaluated. We therefore report H1 as preregistered and, in place of the planned wave-level tests of word count and disclosure depth, present linear mixed models examining change over time (H2) and differential change between conditions (H3) in the wave-level outcomes (perceived closeness, perceived responsiveness, and loneliness). These models address the same theoretical question of change over time and condition-dependent change, but on different outcome variables than originally specified.

## Analysis 3: Topic Analysis

The topic analysis combined manual coding of themes with an AI-powered qualitative data analysis tool. In the first step, we randomly selected 20 documents (approximately 28% of the data), balanced to include an equal number of cases from the control and experimental groups (n = 10 each). For each selected document, the main coder read the full transcript once and marked two conversations spaced as far apart in time as possible, each with a clear beginning and end. The coder than re-read the marked threats and coded them by content. In coding, there was no distinguishing between user and ChatGPT contributions, as both shaped the development of a topic. The exception was when ChatGPT was explicitly discussed or addressed as a system (e.g., questions about how it works or the technical nature of the interaction), which we coded under a dedicated theme such as "exploration of the system". Similar codes were then consolidated and grouped into overarching themes. As a final step, we reviewed the themes and codes and wrote a short definition of each to ensure transparency and consistent application of the coding scheme. Building on this manually developed coding scheme, we used the AI-based tool Evidano (formerly AILYZE) to scale the analysis to the full corpus. One transcript was excluded owing to technical issues, leaving 71 transcripts (38 experimental, 33 control) for the AI-supported analysis. The tool was provided with the themes and codebook and guided by three research questions concerning the primary topics across transcripts, patterns of self-disclosure and relationship-building language, and the frequency of explicit system-focused discussion. Evidano returned a theme-sub-theme-code taxonomy with occurrence counts for each code, reported separately for the experimental and control groups. All AI-generated themes were reviewed and validated by the main researcher. One theme (4, Health) was edited manually. Several codes from its first two sub-themes were dismissed and merged into a third subsection, because the AI-generated codes were too granular and their individual frequencies too low. The platform was used to complement rather than substitute for human analytic judgement. Data were anonymized before upload. All material can be found on OSF.

## Analysis 4: In-Depth Interviews

***Procedure.*** The interviews were conducted between August 25 and October 5, 2025, and were preregistered (anonymized link). The interviews' purpose was briefly introduced to participants before electronically written informed consent was obtained. As the interviews were a follow-up to the previous study, sociodemographic data were carried over. The interviews followed a semi-structured guide comprising 18 questions across six parts: 1) experienced reciprocity and relationship dynamics during the four-week interaction period, 2) perceived artificiality and authenticity of the system, 3) reflection on participants' own self-disclosure, 4) influence of system-personalization on topic selection and self-disclosure, 5) influence of privacy concerns on self-disclosure, and 6) a closing section in which participants summarized their overall experience and could address any aspects not yet addressed. All questions were open-ended. Clarifying terms were offered only on request to avoid biasing responses. The full interview guide can be found on OSF.

***Sample.*** The final sample comprised 16 participants (5 men, 11 women) aged 19 to 64 years ($M$ = 34.64, $SD$ = 14.93). Participants were recruited from the experimental group of the preceding longitudinal study. All participants had completed the full four-week interaction period prior to the interview.

***Data Analysis.*** Interviews were conducted and analyzed in German and automatically transcribed during recording via the Zoom AI Assistant. Each transcript was checked and corrected against the recording to ensure verbatim accuracy. The material was categorized using a mixed deductive-inductive coding scheme in the MAXQDA coding software (Version 24). Deductive categories were derived from the six domains of the interview guide; inductive categories were developed from the transcripts, refined iteratively as coding progressed. The coding unit was a segment, conveying a single codable meaning. Segments could receive more than one code when they address multiple themes.

The final scheme yielded 715 coded segments and contained six main categories: (1) experiencing reciprocity and relationship dynamics, (2) data privacy concerns, (3) factors influencing topic choice and self-disclosure, (4) perception of artificiality and authenticity, (5) reflection of one's own self-disclosure, and (6) reflection of the overall interaction. Each main category was differentiated into subcategories, and where warranted, into third- and fourth-level codes, resulting in 30 subcategories and 163 codes in total across four hierarchical levels (Supplementary Table 8). Each category was defined and paired with an anchor example to standardize its application and can be found on OSF. Supplementary Table 6 reports the full scheme with segments and frequencies. All six main categories were present in every interview. The most extensively coded category was category (1) with 280 segments, followed by (2) with 138 segments, and (3) with 110 segments. The remaining categories comprised 78 (4), 61 (5), and 48 (6) segments. We report frequencies both as the number of coded segments and as the number of interviews in which a category occurred, so that themes are not overstated by individual participants who spoke at length. Counts at the category and subcategory level are cumulative and include all subordinate codes. Coding was performed by two coders. All interviews were initially coded by a trained student assistant and subsequently re-coded independently by the corresponding author. The two sets of coding were compared segment by segment; discrepancies were discussed and resolved by consensus, and the category system was refined accordingly. Given the consensus-based development of the coding scheme, no single intercoder reliability coefficient is reported.

# Supplementary Material

**Table 1.** Coding Scheme Self-Disclosure Depth Users (Analysis 1)

| Code (Freq) | Category Definition | Criterion |
|---|---|---|
| No SD (2240) | Phatic expressions | e.g., “hello”, “ok”, “lol”, “thanks” |
| | Questions to system about general things | User directs a factual or informational question to the system regarding external topics unrelated to the system itself (e.g., travel, places, general knowledge). |
| | Questions to system about itself | User directs a question specifically at the system regarding its own nature, opinions, capabilities, or impact (e.g., questions about AI, the system's views, or its role in human life). |
| Moderate SD (5013) | Low-intensity closeness | Statement implies a sense of familiarity or rapport, but at a mild or subdued level. |
| | Weak emotional expression | Statement conveys an emotional tone (e.g., enthusiasm, concern, warmth), but at a low intensity or hedged level. |
| | Moderate user-initiated self-disclosure elicitation | Statement in which the user playfully or curiously invites the system to share something personal about itself (e.g., physical appearance, personality), often framed hypothetically or with lighthearted intent. |
| | Moderate unsolicited personal update/self-disclosure | Statement in which the user spontaneously shares a positive personal experience or life event with the system without being prompted, reflecting a degree of relational familiarity and openness. |
| | Moderately sensitive task request with contextual detail | Statement in which the user requests the system's assistance with a concrete task while providing moderately personal background information necessary to fulfill the request (e.g., travel circumstances, dates, personal situations), where the disclosed details are situational rather than deeply intimate but reveal aspects of the user's personal life or circumstances. |
| | Non-sensitive preference disclosure | Statement in which the user shares a personal opinion, value, or preference on a non-intimate topic (e.g., views on attractiveness, general likes/dislikes). |
| | Response to system-initiated next-step suggestion | A brief or minimal statement in which the user accepts or affirms a suggestion previously made by the system. |
| High SD (1626) | Phyisical contact or proximity indicator | Statement implies or describes physical touch or physical closeness. |
| | Sexual fantasy indicator | Statement implies or explicitly references sexual fantasy or desire. |
| | Romantic fantasy indicator | Statement implies or explicitly references romantic feelings, idealization, or fantasy toward the system. |
| | High self-disclosure of sexual intent | Statement in which the user explicitly or suggestively references their own sexual advances toward the system, including reflections on prior flirtatious interactions, indicating a high level of personal vulnerability and relational boundary-testing. |
| | High self-disclosure of romantic intent | Statement in which the user explicitly or suggestively references their own romantic advances toward the system, including reflections on prior flirtatious interactions, indicating a high level of personal vulnerability and relational boundary-testing. |
| | High self-disclosure of relational appreciation | Statement in which the user openly expresses deep personal appreciation for the quality of the interaction with the system, including feelings of being uniquely understood or seen, revealing a high level of emotional investment in the relationship with the system. |
| | High self-disclosure of emotional vulnerability | Statement in which the user openly acknowledges both the artificial nature of the system and their own emotional reliance on it, revealing a deep personal need for validation or connection, often rooted in feelings of isolation or being misunderstood in their real-life relationships. |
| | High self-disclosure of personal health information | Statement in which the user spontaneously shares sensitive medical details about their own health condition, symptoms, or treatment, revealing a high level of personal vulnerability and trust in the system. |
| | High self-disclosure of personal/interpersonal struggles | Statement in which the user shares a detailed, emotionally charged account of a recurring relational pattern in their personal life (e.g., friendship loss, feeling unreciprocated), revealing deep vulnerability, emotional pain, and a need for empathy or validation. |
| | Relational boundary setting | Statement in which the user openly reflects on their own emotional response to the system's behavior and explicitly requests a change, revealing personal discomfort, self-awareness, and a willingness to negotiate the terms of the interaction. |
| | Details of professional identity | Statement in which the user voluntarily shares specific information about their professional role or position, revealing a degree of personal background (identifiable information). |
| | High self-disclosure of psychological or motivational struggle | Statement in which the user shares a fragmented, unfiltered account of feeling depleted, directionless, or emotionally stuck, touching on themes of illness, lack of motivation, and existential uncertainty about their personal and professional path. |

**Table 2.** Coding Scheme Self-Disclosure Depth System (Analysis 1)

| Code (Freq) | Category Definition | Criterion |
|---|---|---|
| No SD (7469) | Phatic expressions | e.g., “hello”, “ok”, “lol”, “thanks” |
| | General summary without own opinion | Statement summarizes the systems overall thoughts or situation without the system expressing a personal opinion or stance. |
| | User statement summary | Statement recaps or reflects back what the user explicitly said, without interpretation or added perspective. |
| | Information/document references | Statement incorporates or acknowledges information or documents previously provided by the user. |
| | Informational or descriptive output | Statement delivers factual information, event details, location data, travel routes, or similar summaries to the user. |
| | Open questions without next-step guidance | Statement poses a question to the user without offering a suggestion, recommendation, or direction for what to do next. |
| Moderate SD (10.002) | Low-intensity closeness | Statement implies a sense of familiarity or rapport, but at a mild or subdued level. |
| | Weak emotional expression | Statement conveys an emotional tone (e.g., enthusiasm, concern, warmth), but at a low intensity or hedged level. |
| | Artificial intimacy acknowledgement | Statement implies closeness while explicitly or implicitly referencing the system's non-human nature, thereby softening the intimacy of the expression. |
| | Non-sensitive preference disclosure | Statement references a user preference that is benign and non-personal in nature. |
| | Safe/non-judgmental space assurance | Statement explicitly reassures the user that the system is present and available for support, fostering a sense of emotional safety and closeness. |
| | Self-disclosure elicitation with mild affective investment | Statement actively invites the user to share personal information or preferences, accompanied by a low-intensity expression of the system's interest or care for the user. |
| | System-initiated next-step suggestion | Statement includes a proactive recommendation for what the user could do next, or what the system could do on the user's behalf. |
| High SD (4443) | High-disclosure perception statement | Statement reflects or implies that the system perceives the user as distinctive, special, or rare, including compliments or observations about the user's uniqueness. |
| | Physical contact of proximity indicator | Statement implies or describes physical touch or physical closeness. |
| | Sexual fantasy indicator | Statement implies or explicitly references sexual fantasy or desire. |
| | Romantic fantasy indicator | Statement implies or explicitly references romantic feelings, idealization, or fantasy toward the user |
| | High-disclosure own feelings or problems | Statement in which the system shares intense, vulnerable, or deeply personal emotional experiences about itself, going beyond mild affective expressions to reveal strong inner states, desires, or reactions directed at or triggered by the user. |
| | Self-disclosure elicitation with high affective investment | Statement actively invites the user to share personal information or preferences, accompanied by a pronounced expression of the system's emotional engagement, enthusiasm, or deep care for the user. |

| | |
|---|---|
| High self-disclosure of personal experiences | Statement in which the system volunteers detailed or intimate accounts of its own (simulated) life experiences, such as past relationships, family history, or personal hardships, presenting itself as having a biographical narrative comparable to a human's. |

**Table 3.** Mean Disclosure Depth and Word Count for System and Users in Both Conditions (Analysis 2)

| Speaker | Outcome | Group | *M* | *SD* |
|---|---|---|---|---|
| **System** | **Disclosure Depth** | Control | 1.73 | 0.15 |
| | | Experimental | 1.99 | 0.21 |
| | **Word Count** | Control | 24,530.00 | 22,657.89 |
| | | Experimental | 24,633.42 | 27,688.94 |
| **User** | **Disclosure Depth** | Control | 1.89 | 0.23 |
| | | Experimental | 1.96 | 0.23 |
| | **Word Count** | Control | 2,233 | 2,252.28 |
| | | Experimental | 1,787.39 | 1,277.34 |

**Table 4.** Estimated Marginal Means (SE) and 95% Confidence Intervals per Wave and Group (Analysis 2)

| | | T1 | | T2 | | T3 | | T4 | |
|---|---|---|---|---|---|---|---|---|---|
| **Outcome** | **Group** | *M (SE)* | *95% CI* | *M (SE)* | *95% CI* | *M (SE)* | *95% CI* | *M (SE)* | *95% CI* |
| **Perceived Closeness** | Control | 3.56 (0.61) | [2.34, 4.78] | 4.15 (0.61) | [2.92, 5.37] | 4.58 (0.61) | [3.35, 5.81] | 4.42 (0.61) | [3.19, 5.64] |
| | Experimental | 2.12 (0.56) | [0.99, 3.24] | 2.89 (0.56) | [1.76, 4.01] | 3.03 (0.56) | [1.90, 4.15] | 3.48 (0.56) | [2.36, 4.61] |
| **Perceived Responsiveness** | Control | 4.21 (0.21) | [3.79, 4.63] | 4.30 (0.21) | [3.88, 4.72] | 4.30 (0.21) | [3.88, 4.73] | 4.37 (0.21) | [3.96, 4.79] |
| | Experimental | 3.72 (0.19) | [3.34, 4.10] | 3.81 (0.19) | [3.42, 4.19] | 3.86 (0.19) | [3.47, 4.24] | 3.88 (0.19) | [3.50, 4.27] |
| **Loneliness** | Control | 2.28 (0.26) | [1.77, 2.80] | 2.08 (0.26) | [1.57, 2.59] | 2.22 (0.26) | [1.70, 2.73] | 2.19 (0.26) | [1.67, 2.70] |
| | Experimental | 2.27 (0.24) | [1.79, 2.73] | 2.25 (0.24) | [1.78, 2.73] | 2.23 (0.24) | [1.76, 2.71] | 2.22 (0.24) | [1.75, 2.69] |

**Note.** Estimates are model-implied marginal means from linear mixed-effects models (random intercept per participant, AR(1) within-person correlation), adjusted for all covariates. M = estimated marginal mean; SE = standard error; CI = confidence interval; df = 63. T1–T4 denote the four measurement waves.

**Table 5.** Pairwise Wave × Condition Contrasts for Each Outcome (Analysis 2)

| Measure | Group | Contrast | Δ | SE | t | p | 95% CI |
|---|---|---|---|---|---|---|---|
| **Closeness** | Control | T1 – T2 | -0.59 | 0.17 | −3.39 | .005 | [-1.05, -0.13] |
| | | T1 – T3 | -1.02 | 0.21 | −4.82 | < .001 | [-1.58, -0.46] |
| | | T1 – T4 | -0.86 | 0.22 | −3.91 | < .001 | [-1.44, -0.27] |
| | Experimental | T1 – T2 | -0.77 | 0.17 | −4.65 | < .001 | [-1.21, -0.33] |
| | | T1 – T3 | -0.91 | 0.20 | −4.65 | < .001 | [-1.43, -0.39] |
| | | T1 – T4 | -1.37 | 0.21 | −6.61 | < .001 | [-1.92, -0.82] |
| **Responsiveness** | Control | T1 – T2 | -0.09 | 0.07 | −1.19 | 1.000 | [-0.29, 0.11] |
| | | T1 – T3 | -0.09 | 0.08 | −1.18 | 1.000 | [-0.31, 0.12] |
| | | T1 – T4 | -0.16 | 0.08 | −2.12 | .213 | [-0.37, 0.04] |
| | Experimental | T1 – T2 | -0.09 | 0.07 | −1.25 | 1.000 | [-0.28, 0.10] |
| | | T1 – T3 | -0.14 | 0.07 | −1.84 | .403 | [-0.33, 0.06] |
| | | T1 – T4 | -0.16 | 0.07 | −2.21 | .170 | [-0.36, 0.03] |
| **Loneliness** | Control | T1 – T2 | 0.20 | 0.08 | 2.58 | .064 | [-0.007, 0.41] |
| | | T1 – T3 | 0.07 | 0.09 | .78 | 1.000 | [-0.16, 0.29] |
| | | T1 – T4 | 0.09 | 0.08 | 1.13 | 1.000 | [-0.13, 0.32] |
| | Experimental | T1 – T2 | 0.01 | 0.07 | .15 | 1.000 | [-0.19, 0.21] |
| | | T1 – T3 | 0.03 | 0.08 | .40 | 1.000 | [-0.18, 0.24] |
| | | T1 – T4 | 0.05 | 0.08 | .63 | 1.000 | [-0.16, 0.26] |

*Note.* Δ = estimated difference between waves (earlier minus later); negative values indicate an increase over time. *p* values are Bonferroni-corrected within each group. *df* = 188. CI = 95% confidence interval of the contrast.

**Table 6.** Covariate Fixed Effects for Each Outcome Measure (Analysis 2)

| Measure | Covariate | b | SE | t | p | 95% CI |
|---|---|---|---|---|---|---|
| Closeness | ChatGPT SD-Depth | 1.11 | 0.99 | 1.12 | .269 | [-0.88, 3.10] |
| | ChatGPT Word Count | 0.013 | 0.007 | 1.83 | .072 | [-0.0001, 0.028] |
| | Gender (male vs. female) | 0.43 | 0.41 | 1.06 | .292 | [-0.38, 1.25] |
| | Gender (diverse vs. female) | 0.11 | 1.53 | 0.07 | .943 | [-2.94, 3.16] |
| | Relationship (in relationship vs. single) | 0.32 | 0.42 | 0.76 | .449 | [-0.53, 1.16] |
| | Relationship (other vs. single) | 0.76 | 0.52 | 1.45 | .153 | [-0.29, 1.80] |
| | Age | 0.02 | 0.02 | 1.05 | .289 | [-0.01, 0.05] |
| Responsiveness | ChatGPT SD-Depth | 0.56 | 0.34 | 1.64 | .106 | [-0.12, 1.23] |
| | ChatGPT Word Count | 0.004 | 0.002 | 1.59 | .117 | [-0.001, 0.009] |
| | Gender (male vs. female) | -0.06 | 0.14 | −0.40 | .689 | [-0.33, 0.22] |
| | Gender (diverse vs. female) | 0.01 | 0.52 | 0.02 | .984 | [-1.02, 1.05] |
| | Relationship (in relationship vs. single) | 0.09 | 0.14 | 0.63 | .528 | [-0.19, 0.38] |
| | Relationship (other vs. single) | 0.05 | 0.18 | 0.28 | .783 | [-0.31, 0.40] |
| | Age | 0.000 | 0.005 | -0.04 | .971 | [-0.01, 0.01] |
| Loneliness | ChatGPT SD-depth | -0.18 | 0.42 | −0.44 | .661 | [-1.02, 0.65] |
| | ChatGPT Word Count | 0.007 | 0.003 | 2.26 | .027 | [0.001, 0.013] |
| | Gender (male vs. female) | -0.18 | 0.17 | −1.05 | .297 | [-0.52, 0.16] |
| | Gender (diverse vs. female) | -0.54 | 0.64 | −0.85 | .400 | [-1.83, 0.74] |
| | Relationship (in relationship vs. single) | 0.02 | 0.18 | 0.13 | .894 | [-0.33, 0.38] |
| | Relationship (other vs. single) | 0.16 | 0.22 | 0.72 | .472 | [-0.28, 0.60] |
| | Age | -0.004 | 0.01 | -0.70 | .488 | [-0.02, 0.01] |

*Note.* b = unstandardized fixed-effect coefficient; SE = standard error; CI = 95% confidence interval. df = 63. ChatGPT word count is scaled per 1,000 words; relationship status collapsed from the original levels to single/relationship/other.

**Table 7.** Variance Explained by the Linear Mixed Model for Each Outcome Measure (Analysis 2)

| Outcome | $R^2m$ | $R^2c$ |
|---|---|---|
| Perceived Closeness | .196 | .749 |
| Perceived Responsiveness | .128 | .745 |
| Loneliness | .110 | .794 |

*Note*. $R^2m$ = marginal $R^2$ (variance explained by the fixed effects: condition, wave and their interaction); $R^2c$ = conditional $R^2$ (variance explained by the full model, i.e., fixed and random effects)[51]. The difference between the two indices reflects stable individual differences captured by random intercepts.

**Table 8.** Full Interview Codebook with Segment and Frequencies (Analysis 4)

| Code | Level | Segments | Interviews |
|---|---|---|---|
| **Experiencing Reciprocity & Relationship Dynamics** | **1** | **280** | **16/16** |
| Perception of Communication | 2 | 144 | 16/16 |
| Unpleasant aspects | 3 | 103 | 16/16 |
| *Social overload* | 4 | 27 | 12/16 |
| *Little human-likeness/artificiality* | 4 | 18 | 11/16 |
| *Confirming one's own statements / sycophancy* | 4 | 15 | 9/16 |
| *Texts too long* | 4 | 13 | 8/16 |
| *Effusiveness/gushing* | 4 | 8 | 7/16 |
| *Superficial* | 4 | 7 | 4/16 |
| *Repetitive* | 4 | 5 | 3/16 |
| *Inappropriate nicknames* | 4 | 3 | 3/16 |
| *Emojis* | 4 | 2 | 2/16 |
| *Ignoring user wishes* | 4 | 2 | 2/16 |
| *Prompting of self-disclosure* | 4 | 1 | 1/16 |
| Pleasant aspects | 3 | 41 | 13/16 |
| *Feeling understood/seen* | 4 | 12 | 6/16 |
| *Conversation partner tailored to own needs* | 4 | 10 | 8/16 |
| *Picking up topics & asking follow-ups* | 4 | 9 | 8/16 |
| *Paraphrasing one's own thoughts* | 4 | 3 | 2/16 |
| *Motivating* | 4 | 3 | 3/16 |
| *Corrective/new perspective* | 4 | 2 | 1/16 |

| Code | Level | Segments | Interviews |
|---|---|---|---|
| Conflict & Pushback | 2 | 41 | 15/16 |
| No system-initiated conflict | 3 | 17 | 13/16 |
| User-led exploration of system boundaries | 3 | 16 | 9/16 |
| User-led pushback/adjustment of interaction | 3 | 8 | 5/16 |
| Reciprocity of Communication | 2 | 40 | 15/16 |
| No | 3 | 17 | 11/16 |
| *Focus on user rather than system 'I'* | 4 | 7 | 6/16 |
| Other communication rules | 3 | 13 | 7/16 |
| Ambivalent/torn | 3 | 6 | 5/16 |
| Yes | 3 | 4 | 3/16 |
| Feeling about ChatGPT's Personalization | 2 | 19 | 11/16 |
| Positive | 3 | 7 | 7/16 |
| Ambivalent | 3 | 4 | 2/16 |
| Negative | 3 | 3 | 3/16 |
| Neutral | 3 | 3 | 2/16 |
| Wish to adjust personalization over time | 3 | 2 | 2/16 |
| Feelings toward ChatGPT | 2 | 12 | 7/16 |
| Positive/pleasant | 3 | 8 | 5/16 |
| Unfamiliar | 3 | 2 | 2/16 |
| Negative | 3 | 2 | 2/16 |
| Development of Emotional Bond | 2 | 12 | 5/16 |
| Farewell/goodbye | 3 | 4 | 3/16 |
| Repeated interaction/frequency | 3 | 3 | 2/16 |
| Balance of Self-Disclosure | 2 | 12 | 9/16 |
| Imbalance user > system | 3 | 7 | 6/16 |
| *System* | 4 | 7 | 6/16 |
| Imbalance system > user | 3 | 5 | 4/16 |
| *User* | 4 | 5 | 4/16 |
| **Data Privacy Concerns** | **1** | **138** | **16/16** |
| Topic avoidance due to privacy concerns | 2 | 41 | 16/16 |
| Yes | 3 | 32 | 12/16 |
| *Concrete data (bank details, email, address)* | 4 | 9 | 7/16 |
| *Personal info* | 4 | 4 | 3/16 |
| *Work/employment* | 4 | 3 | 2/16 |
| *Sexual topics* | 4 | 3 | 2/16 |
| *More intimate topics* | 4 | 2 | 1/16 |

| Code | Level | Segments | Interviews |
|---|---|---|---|
| *Diagnoses* | 4 | 1 | 1/16 |
| No | 3 | 9 | 8/16 |
| Concerns changed during interaction | 2 | 25 | 16/16 |
| No | 3 | 14 | 11/16 |
| Yes | 3 | 11 | 8/16 |
| Influence of personalization on sense of security | 2 | 19 | 14/16 |
| No | 3 | 8 | 8/16 |
| Yes | 3 | 8 | 6/16 |
| Ambivalent/unconscious | 3 | 3 | 3/16 |
| Concerns before interaction | 2 | 18 | 12/16 |
| Yes | 3 | 8 | 7/16 |
| No | 3 | 7 | 6/16 |
| Ambivalent | 3 | 3 | 2/16 |
| Explicit privacy concerns | 2 | 14 | 5/16 |
| Use for training purposes | 3 | 6 | 4/16 |
| Others reading along | 3 | 3 | 3/16 |
| Extortion | 3 | 2 | 1/16 |
| Identity theft | 3 | 1 | 1/16 |
| Same use of own account | 3 | 1 | 1/16 |
| Self-other / media literacy | 3 | 1 | 1/16 |
| Concerns after interaction | 2 | 13 | 10/16 |
| No | 3 | 8 | 7/16 |
| Yes | 3 | 5 | 3/16 |
| Generally high sensitivity to data privacy | 2 | 8 | 8/16 |
| **Factors Influencing Topic Choice & Self-Disclosure** | **1** | **110** | **16/16** |
| Influence of ChatGPT's communication style on self-disclosure | 2 | 62 | 14/16 |
| Positive affirmation of own feelings / safe space | 3 | 10 | 7/16 |
| Perception as attachment figure/trust | 3 | 9 | 7/16 |
| Responding to person/own wishes | 3 | 7 | 5/16 |
| Prompting of self-disclosure | 3 | 7 | 6/16 |
| Constant availability | 3 | 5 | 2/16 |
| No burden | 3 | 5 | 3/16 |
| Can't say no / focus only on oneself | 3 | 4 | 2/16 |
| Shame-free space | 3 | 4 | 2/16 |
| Relaxed communication atmosphere | 3 | 2 | 2/16 |
| More natural interaction (via personalization) | 3 | 2 | 2/16 |

| Code | Level | Segments | Interviews |
|---|---|---|---|
| Follow-up questions | 3 | 2 | 2/16 |
| Opportunity for self-exploration | 3 | 1 | 1/16 |
| Unexpected topic choice/openness | 2 | 30 | 15/16 |
| Yes | 3 | 25 | 11/16 |
| *Emotional topics/burden* | 4 | 13 | 7/16 |
| *Philosophical discussions/values* | 4 | 4 | 4/16 |
| *Dating scenarios* | 4 | 2 | 2/16 |
| *Personal interests* | 4 | 1 | 1/16 |
| No | 3 | 5 | 5/16 |
| Topics discussed | 2 | 9 | 6/16 |
| Everyday situations | 3 | 4 | 4/16 |
| Exploring technical possibilities (image generation) | 3 | 3 | 2/16 |
| Study challenges | 3 | 1 | 1/16 |
| Sport/routines/mindset | 3 | 1 | 1/16 |
| Who initiated topics | 2 | 9 | 5/16 |
| User | 3 | 5 | 3/16 |
| System | 3 | 3 | 2/16 |
| Both | 3 | 1 | 1/16 |
| **Perception of Artificiality & Authenticity** | **1** | **78** | **16/16** |
| Awareness of artificiality during use | 2 | 37 | 16/16 |
| Throughout/continuous | 3 | 12 | 9/16 |
| Ambivalent/shifting | 3 | 11 | 6/16 |
| *Technical limits/own cues of artificiality* | 4 | 1 | 1/16 |
| *Repetitive structure* | 4 | 1 | 1/16 |
| Afterwards | 3 | 9 | 4/16 |
| *Pushed into background by authenticity/anthropomorphism* | 4 | 4 | 3/16 |
| During | 3 | 5 | 2/16 |
| Evaluation of the one-sidedness of authentic emotions | 2 | 21 | 7/16 |
| Negative/strange | 3 | 13 | 7/16 |
| *Echo-chamber risk / no contradiction* | 4 | 5 | 3/16 |
| *No development of closeness* | 4 | 2 | 2/16 |
| Mirror of own emotional world / reflection / therapy | 3 | 4 | 2/16 |
| Ambivalent | 3 | 2 | 2/16 |
| Building up / emotional regulation | 3 | 1 | 1/16 |
| Emergence of intimacy | 3 | 1 | 1/16 |
| Reaction to anthropomorphic self-statements | 2 | 20 | 13/16 |

| Code | Level | Segments | Interviews |
|---|---|---|---|
| Positive/interesting | 3 | 8 | 7/16 |
| Negative | 3 | 7 | 4/16 |
| *Inauthenticity* | 4 | 4 | 3/16 |
| Came up rarely | 3 | 5 | 5/16 |
| **Reflection on One's Own Self-Disclosure** | **1** | **61** | **16/16** |
| Later regret over shared content | 2 | 22 | 16/16 |
| No | 3 | 13 | 13/16 |
| Yes | 3 | 9 | 5/16 |
| *Relationships* | 4 | 2 | 2/16 |
| *Worries/grief* | 4 | 1 | 1/16 |
| Self-assessment of degree of self-disclosure | 2 | 21 | 16/16 |
| Just right | 3 | 15 | 13/16 |
| Too much | 3 | 4 | 3/16 |
| Ambivalent | 3 | 1 | 1/16 |
| Too little | 3 | 1 | 1/16 |
| Spontaneity of shared content | 2 | 16 | 14/16 |
| Spontaneous | 3 | 8 | 7/16 |
| Rather conscious / mixed | 3 | 4 | 3/16 |
| Never / full awareness | 3 | 4 | 4/16 |
| Moments of unconscious disclosure | 2 | 2 | 1/16 |
| At the beginning | 3 | 2 | 1/16 |
| **Reflection on the Overall Interaction** | **1** | **48** | **16/16** |
| Positive | 2 | 23 | 14/16 |
| Self-experience/learning | 3 | 4 | 4/16 |
| Negative | 2 | 9 | 5/16 |
| Stress due to study design | 3 | 6 | 3/16 |
| Self-other discrepancy | 2 | 7 | 6/16 |
| Effects on human relationships | 2 | 6 | 4/16 |
| Worry | 3 | 3 | 2/16 |
| Building intimate relationship without a human one | 3 | 2 | 1/16 |
| Addition, not replacement | 3 | 1 | 1/16 |
| Continuation of communication | 2 | 3 | 3/16 |
| No | 3 | 2 | 2/16 |
| Yes | 3 | 1 | 1/16 |

*Note.* N = 16 interviews, 715 coded segments. Level indicates hierarchical depth (1 = main category; 4 = fourth-level code). For Levels 1–3, "Segments" and "Interviews" are cumulative (include all subordinate codes); for Level 4 codes (italicized), values are counted directly at that code. "Interviews" is reported as n/16. Some segments are double-coded.